\documentclass[prd,aps,preprintnumbers, showpacs, nofootinbib,superscriptaddress,
notitlepage
]{revtex4-1}

\usepackage{epsfig}
\usepackage{bm}
\usepackage{amssymb}
\usepackage{amsmath}
\usepackage{color}
\usepackage{subfigure}
\usepackage{slashed}
\usepackage{tikz}
\newcommand*\circled[1]{\tikz[baseline=(char.base)]{
  \node[shape=circle,draw,inner sep=0.5pt] (char) {#1};}}

\newcommand{\be}{\begin{equation}}
\newcommand{\ee}{\end{equation}}
\newcommand{\bea}{\begin{eqnarray}}
\newcommand{\eea}{\end{eqnarray}}

\begin{document}
\title{Medium Induced Transverse Momentum Broadening in Hard Processes}

\author{A. H. Mueller}
\affiliation{Department of Physics, Columbia University, New York, NY 10027, USA}

\author{Bin Wu}
\affiliation{Department of Physics, The Ohio State University, Columbus, OH 43210, USA}
\affiliation{Institut de Physique Theorique, CEA Saclay, UMR 3681, F-91191 Gif-sur-Yvette, France}

\author{Bo-Wen Xiao}
\affiliation{Key Laboratory of Quark and Lepton Physics (MOE) and Institute
of Particle Physics, Central China Normal University, Wuhan 430079, China}

\author{Feng Yuan}
\affiliation{Nuclear Science Division, Lawrence Berkeley National
Laboratory, Berkeley, CA 94720, USA}

\begin{abstract}
Using deep inelastic scattering on a large nucleus as an example, we consider the transverse momentum broadening of partons in hard processes in the presence of medium. We find that one can factorize the vacuum radiation contribution and medium related $P_T$ broadening effects into the Sudakov factor and medium dependent distributions, respectively. Our derivations can be generalized to other hard processes, such as dijet productions, which can be used as a probe to measure the medium $P_T$ broadening effects in heavy ion collisions when Sudakov effects are not overwhelming. \end{abstract}
\maketitle

\section{Introduction}

One of the most intriguing discoveries at the Relativistic Heavy Ion Collider (RHIC) is the strongly coupled quark gluon plasma (QGP)\cite{Gyulassy:2004zy} created in the heavy ion collisions. There have been great experimental efforts on the quantitative study of various properties of QGP in terms of both energy loss and transverse momentum $P_T$ broadening effects\cite{Gyulassy:1993hr, Baier:1996kr, Baier:1996sk, Baier:1998kq, Zakharov:1996fv}. For example, as a clear indication of a jet quenching effect due to large energy loss, a large suppression of the single hadron spectra in the high $P_T$ region in central $AuAu$ collisions has been observed\cite{Adams:2003kv, Adams:2003im, Adler:2003qi}. In addition, RHIC\cite{Adler:2002tq} has also observed that the back-to-back hadron correlations for moderate $P_T$ disappear for central $AuAu$ collisions. Although one can attribute this effect to both energy loss and $P_T$ broadening effects, it is believed that the normalized angular correlation around $\Delta \phi \sim \pi$ is mostly due to medium transverse momentum broadening with $\Delta \phi$ being the azimuthal angle difference between the trigger hadron and the associate hadron. 

In fact, it was shown in the Baier-Dokshitzer-Mueller-Peigne-Schiff (BDMPS) approach\cite{Baier:1996kr, Baier:1996sk, Baier:1998kq} that the energy loss and $P_T$ broadening effects are related through the following formula $-\frac{dE}{dx}\simeq \frac{\alpha_s N_c}{4} \hat{q} L $, where $\hat{q} L$ represents the typical transverse momentum squared that a parton acquires in the medium of length $L$ . Here, $\hat q$ is the so-called jet-quenching parameter which depends on the density of the QGP medium. Therefore, one would expect that the energy loss effect should be tied together with the transverse momentum broadening effects in heavy ion experiments. 

Since the commencement of the LHC, similar suppression of single hadron spectra\cite{CMS:2012aa, Abelev:2012hxa} and inclusive jets \cite{Aad:2012vca} has also been found in $PbPb$ collisions, which implies that similar jet quenching effects persist in the LHC regime. In the meantime, approximately a factor of two suppression of the back-to-back dihadron correlation with $ 8 \, \textrm{GeV}<P_{T, trig} <15 \, \textrm{GeV}$\cite{Aamodt:2011vg} in central heavy ion collisions also suggests the presence of significant medium effects.  

Nevertheless, the dijet measurements conducted by CMS and ATLAS at the LHC\cite{Chatrchyan:2011sx, Aad:2010bu} seem to be a bit puzzling at first sight. On one hand, they observed striking dijet asymmetries in central $PbPb$ collisions which is consistent with the jet quenching effect\cite{Qin:2010mn}. Since the dijet asymmetry strongly depends on the transverse energy difference of the dijet system, this observable is not as sensitive to the $P_T$ broadening of jets as the angular correlation. On the other hand, there is no trace of significant angular decorrelation found in the same dijet measurement. As a matter of fact, the normliazed angular distribution in central $PbPb$ collisions is almost the same as the one measured in $pp$ collisions for $\Delta \phi > 2$.

From the theoretical point of view, there are mainly two competing contributions to the correlation (decorrelation) of the dijet angular distribution in high energy heavy ion collisions, namely, the Sudakov effect and the medium induced $P_T$ broadening (For the normalized angular distribution as shown in Ref.~\cite{Chatrchyan:2011sx}, one expects that the energy loss effect is not very important.). The Sudakov effect, also known as the parton shower, has been an important topic of QCD studies for several decades. It normally occurs due to large amounts of gluon radiation in hard processes, such as high invariant mass Drell-Yan lepton pair production process as well as the $W$ and $Z$ boson production\cite{Collins:1984kg}. Especially, recent studies\cite{Banfi:2008qs, Mueller:2012uf, Mueller:2013wwa, Sun:2014gfa} in several areas of QCD have shown that it is important to perform the Sudakov resummation in order to obtain a consistent description of back-to-back dijet angular correlations in hard processes. It is also important to mention that the Sudakov factor arises from the incomplete cancellation of real and virtual graphs in high order perturbative calculations if we are measuring the transverse momentum of the high mass Drell-Yan lepton pair (or the transverse momentum of heavy particles) or the momentum imbalance (or the angular correlation) of dijets produced in high energy scattering. If one integrates over the transverse momentum of the produced particle or the azimuthal angle difference of dijets, the Sudakov effect disappears since the real-virtual cancellation becomes more complete after the integration. 

In order to quantitatively study $P_T$ broadening effects in back-to-back dijet angular correlation measurements with the presence of medium effects, we need to develop a sophisticated formalism which incorporates Sudakov effects and the medium induced $P_T$ broadening effects, and investigate the interplay of these two effects in different experimental environments. 
In general, one expects that the medium effects are absent in $pp$ collisions, and the correlations are solely due to Sudakov effects in the back-to-back dijet configurations. This has led to the successful description\cite{Sun:2014gfa} of the Tevatron ($p\bar p$)\cite{Abazov:2004hm} and the LHC ($pp$)\cite{Khachatryan:2011zj} dijet correlation data. Generally speaking, the larger the collision energy and jet transverse momentum are, the larger the Sudakov effects are. In the case of $pA$\cite{Chatrchyan:2014hqa} and $AA$\cite{Aad:2010bu, Chatrchyan:2011sx} collisions, the produced dijet system can also interact with either the cold nuclear medium or the hot-dense QGP medium, which generates extra transverse momentum broadening effects. In dijet productions at the LHC with the transverse momentum of the leading jet larger than $100 \, \textrm{GeV}$, Sudakov effects dominate over medium effects. Rough estimates give the transverse momentum broadening of the Sudakov effect at the LHC energy for dijet productions with $P_T \sim 100\, \textrm{GeV}$ as $\langle \triangle P_T^2\rangle  \sim 100 \, \textrm{GeV}^2$~\cite{Sun:2014gfa}, as opposed to that due to medium effects which is $\langle \triangle P_T^2\rangle  \sim \hat q L \sim 10 \, \textrm{GeV}^2$. Note that since the nature of momentum broadening in the transverse direction is the same as a random walk or Brownian motion, which suggests that we should always compare $\langle \triangle P_T^2\rangle$ instead of $\langle |\triangle P_T|\rangle$. This naturally explains why there are no visible medium modifications found for dijet angular correlation measurement in both $pPb$\cite{Chatrchyan:2014hqa} and $PbPb$\cite{Aad:2010bu, Chatrchyan:2011sx} collisions at the LHC, since the corresponding modification in terms of dijet angular distributions is too small to be seen at the LHC. To probe the medium effects through angular correlation measurements, we either need to lower the $P_T$ of the dijet system or measure dihadrons with much lower $P_T$ as in Ref.~\cite{Adler:2002tq, Aamodt:2011vg}. This can significantly reduce the he Sudakov effects. Therefore, as recently pointed out in Ref.~\cite{Mueller:2016gko, Chen:2016vem}, one can also measure medium effects at RHIC through dijets with roughly $P_T \sim 35 \, \textrm{GeV}$ and hadron-jet as well as dihadron correlations.

In this paper we study the transverse momentum distribution of jets produced by a hard scattering in the medium. For explicitness we consider a jet to be produced in the deep inelastic scattering of a transverse virtual photon on a nucleus. We consider in detail two separate cases where (i) the time scale over which the jet is produced, $\tau_q$, is much less than the size, $L$, of the nucleus and (ii) where $\tau_q$ is much greater than $L$ in the target rest frame. The transverse momentum of the jet then comes from various sources, namely, from the hard scattering itself, from radiation not induced by the medium (Sudakov radiation), from multiple scattering of the jet in the medium ($\hat q$) and from radiation induced by the medium (radiative corrections to $\hat q$). In our current discussion we take the transverse momentum of the virtual photon to be zero to minimize the transverse momentum coming from the hard scattering. 

Although our discussion is done in the context of cold nuclear matter, a large nucleus, it is straightforward to extend to hot matter simply by changing from the $\hat q$ of cold matter to the $\hat q$ of hot matter. For example the discussion given in Sec.~\ref{lm}, for $\tau_q \ll L$, can be used to describe the imbalance between the transverse momentum of the two jets produced in a hard scattering in heavy ion collisions. 

In Ref.~\cite{Mueller:2016gko, Chen:2016vem}, the relative importance to imbalance (the azimuthal angle between the two jets, hadron-jet or dihadrons) of Sudakov emission and medium induced broadening (multiple scattering effects together with medium induced radiation) was analyzed for jets produced in heavy ion collisions. In Sec.~\ref{lm} we include the medium induced radiative contribution, namely, radiative corrections to $\hat q$, to the imbalance. If the $\hat q$ of Sec.~\ref{lm} is taken to be that of hot matter then we have evaluated all the contributions to $\hat q$ included in the analysis of Ref.~\cite{Mueller:2016gko, Chen:2016vem}. 

In the case that the transverse momentum broadening is dominated by Sudakov double logarithmic radiation, as in the case of jet production in LHC heavy ion collisions, it is necessary to revisit the evaluation of radiative corrections to $\hat q$ as done in the context of a $\hat q$-dominated broadening. This is done in Sec.~\ref{rcq} where all double logarithmic radiative corrections to $\hat q$ are evaluated. 

In Sec.~\ref{smc} we consider small-$x$ deep inelastic scattering where the jet is formed on a time scale long compared to the length of the medium. We begin in Sec.~\ref{fixed} by doing the analysis assuming a fixed coupling and with the photon virtuality in the scaling region of the small-$x$ evolution. Up to an overall constant we are able to get analytic expressions for the jet broadening in (\ref{e37}) or, in the various regions shown in Fig.~\ref{f6}, in (\ref{e39})-(\ref{e41}). 
It is interesting to investigate what happens at a fixed amount of the broadening, $k_\perp$, of the jet as one varies the hardness, $Q^2$, from moderate to large values while always assuming that $x$ is small enough that one remains in the scaling region of the small-$x$ evolution. When $\ln \frac{Q^2}{k_\perp^2} <\frac{1}{\sqrt{\alpha_s}}$ Sudakov effects are not visible and the transverse momentum, $k_\perp$, comes completely from small-$x$ evolution and exhibits scaling in (\ref{e39}). As $Q^2$ is increased one gets a scaling behavior with a simple factor giving the Sudakov contribution given by (\ref{e40}). When $\ln \frac{Q^2}{k_\perp^2} > \frac{1}{\alpha_s}$ the scaling behavior is completely destroyed and the transverse momentum distribution is flat reflecting the randomizing effects of the Sudakov radiation. This is exhibited in (\ref{e41}). 

In Sec.~\ref{rcs}, running coupling effects are introduced and we no longer suppose that $Q^2$ lie in the scaling region of the small-$x$ evolution. The three different regions of Fig.~\ref{f6} give very similar results as compared to the fixed coupling case. The first region, where $Q^2$ is not so large, shows no Sudakov modification of the spectrum of transverse momentum broadening. The next region of somewhat larger $Q^2$ again has a simple Sudakov factor (see (\ref{e49})) modifying the small-$x$ answer. Finally, the large $Q^2$ region again completely eliminates all $k_\perp$-dependence, as given in (\ref{e56}). 

In the case where $\tau_q \gg L$, $\hat q $-effects are not very visible, since they are hidden in the initial distribution for the small-$x$ evolution. In Sec.~\ref{meq} we show explicitly how $\hat q$-effects, and radiative corrections to $\hat q$, come into the initial condition for small-$x$ evolution. If there were no radiative corrections to $\hat q$, the initial condition for small-$x$ evolution is just the scattering matrix for a dipole given by the McLerran-Venugopalan model. If one uses $\hat{q}_t$ rather than $\hat q$ in the MV model initial condition then evolution in the medium is also included and will show up as an enhancement of $Q_s^2$. We conclude and summarize in Sec.~\ref{con}.

\section{Large medium forward jet production in DIS}
\label{lm}

\subsection{The basic formulas}
We begin our discussions of forward jet production in deep inelastic scattering (DIS) on a large nucleus in the case $\tau_q =\frac{2q_+}{Q^2}$ is much less than the length of the medium. For a scattering at impact parameter $b$ in the nucleus, the nuclear medium length is $L=2\sqrt{R^2-b^2}$ with $R$ the nuclear radius. The (transverse) virtual photon initiating the process has momentum $q_\mu$ with $q_\mu =(q_+, q_-= -\frac{Q^2}{2q_+}, q_\perp=0)$. The process is illustrated in Fig.~\ref{f1} where the forward quark (or antiquark) has momentum $k$ and travels a distance $z$ in the medium after its production. In the current situation of $\tau_q/L \ll 1$, this production can take
place on a definite nucleon in the nucleus with that nucleon at a distance $L-z$ from the front face of the nucleus. 
In Refs.~\cite{Luo:1993ui, Zhang:2014dya,Kang:2016ron}, similar process has been considered to study the modification of average transverse momentum squared due to the medium effects. In this paper, we focus on the transverse momentum spectrum, where all the relevant QCD dynamics
play important roles.

In this large-$x$ process there is no small-$x$ evolution. However, there is the DGLAP evolution of the quark distribution of the struck nucleon,
the Sudakov effects due to the hard scattering and the measurement of the forward quark, and finally the multiple scattering and medium induced radiation of the outgoing quark. At the moment we do not introduce a cone condition for the produced quark jet nor do we consider the fragmentation of the quark. These can be included accordingly for a complete evaluation of the forward jet electro-production. Our purpose here is to illustrate in a simple context the various effects that may occur in jet production in a medium.

The transverse momentum spectrum of the quark is given by
\begin{equation}
\frac{dN}{d^2 b d^2 k_\perp}= \int \frac{d^2 x_\perp}{(2\pi)^2} e^{-ik_\perp\cdot x_\perp} \rho\, xq_N\left(x, \frac{1}{x_\perp^2 +1/Q^2}\right) \int_0^L dz e^{-\mathcal{E}}, \label{spectrum}
\end{equation}
where 
\begin{equation}
\mathcal{E}=\hat{q} x_\perp^2 z/4 +\mathcal{E}_\textrm{Sud}+\mathcal{E}_\textrm{Medium Induced Radiation (MIR)}, \label{tot}
\end{equation}
with the quark transport coefficient 
\begin{equation}
\hat {q} =\frac{C_F}{N_c} \frac{4\pi^2 \alpha_s N_c}{N_c^2-1} \rho \, xG(x) \ . \label{qhatdef}
\end{equation}
Here, $\rho$ is the nucleon density and $xG$ the nucleon's gluon distribution, while $xq_N$ the quark distribution of a nucleon should be evaluated at a scale $x_\perp^2$, that is $q_N=q_N \left(x, \frac{1}{x_\perp^2 +1/Q^2}\right)$. When $x_\perp=0$, see below, one gets $q_N$ as the quark distribution at the hard scattering scale. The various terms of Eq.~(\ref{tot}) can be interpreted as follows: $\hat q$ term accounts for multiple scattering as the quark passes through the nucleus; $\mathcal{E}_\textrm{Sud}$ accounts for the real and virtual Sudakov corrections, which are medium independent, induced by the hard scattering; and $\mathcal{E}_\textrm{MIR}$ accounts for gluonic radiative corrections which involve a single scattering in the medium. As we shall see below, the $\hat q$ and $\mathcal{E}_\textrm{MIR}$ terms in (\ref{tot}) can be combined into a more complete $\hat q$, which we shall call $\hat {q}_t \equiv \hat q _{\textrm{total}}$, where
\begin{equation}
\hat {q}_t x_\perp^2 z/4 =\hat {q} x_\perp^2 z/4  +\mathcal{E}_\textrm{MIR}. 
\end{equation}
Then the $z$-integral in (\ref{spectrum}) can be done giving 
\begin{equation}
\frac{dN}{d^2 b d^2 k_\perp}=\int \frac{d^2 x_\perp}{x_\perp^2} \frac{\rho\, xq_N\left(x, \frac{1}{x_\perp^2 +1/Q^2}\right)}{\pi^2 \hat {q}_t} e^{-ik_\perp\cdot x_\perp} \left(1-e^{-\hat{q}_t x_\perp^2L/4}\right) e^{-\mathcal{E}_{\textrm{Sud}}}. \label{wwquark}
\end{equation}
The right hand side of (\ref{wwquark}) has the form of an unintegrated Weizsacker-Williams quark distribution in analogy with the Weizsacker-Williams (WW) \cite{Kovchegov:1996ty, Kovchegov:1998bi, Kharzeev:2003wz} gluon distribution. We note that 
\begin{equation}
\int \frac{dN}{d^2 b d^2 k_\perp} d^2 b d^2 k_\perp =A xq_N.
\end{equation}
The Sudakov factor in (\ref{wwquark}) is naturally included as part of the WW quark distribution since the usual Wilson line of the WW distribution implicitly includes the Sudakov factor, see the discussions below.

\subsection{The Sudakov factor}
\begin{figure}[tbp]
\begin{center}
\includegraphics[width=12cm]{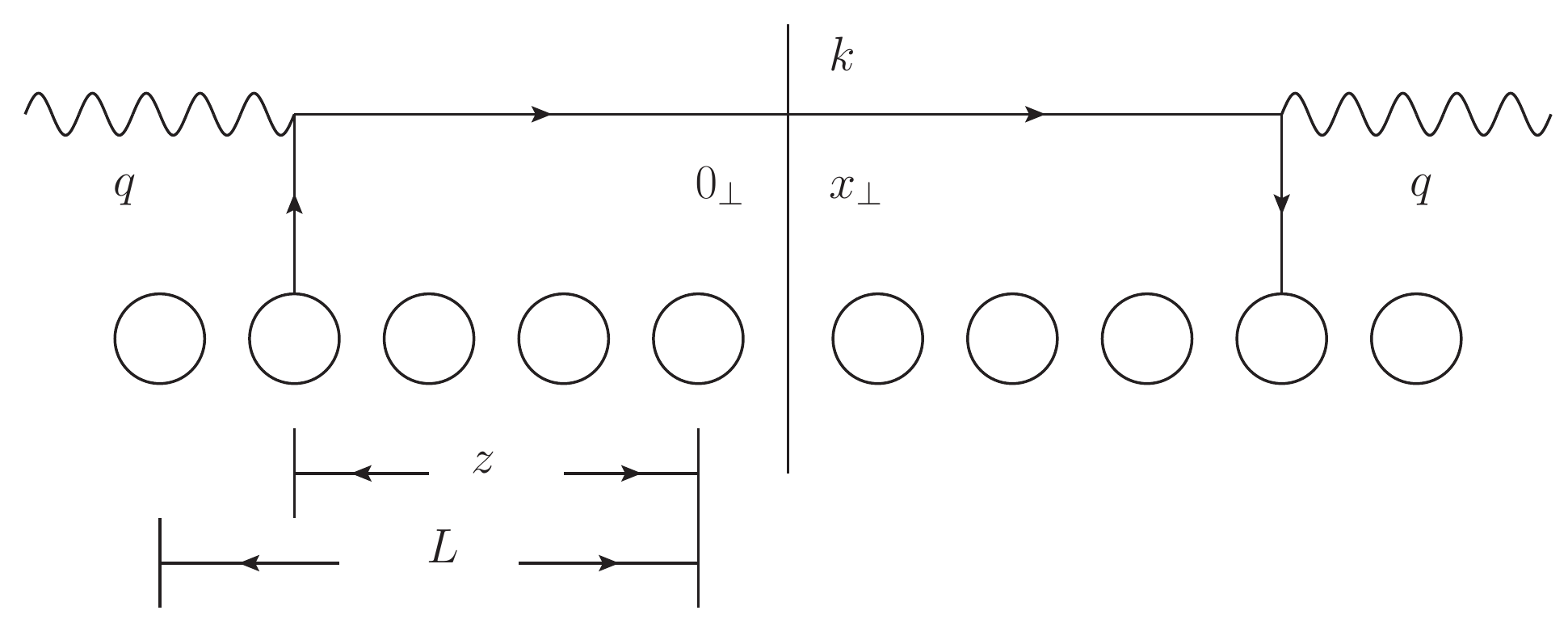}
\end{center}
\caption[*]{Forward jet production in DIS on a large nucleus in the large-$x$ region.}
\label{f1}
\end{figure}

\begin{figure}[tbp]
\begin{center}
\includegraphics[width=12cm]{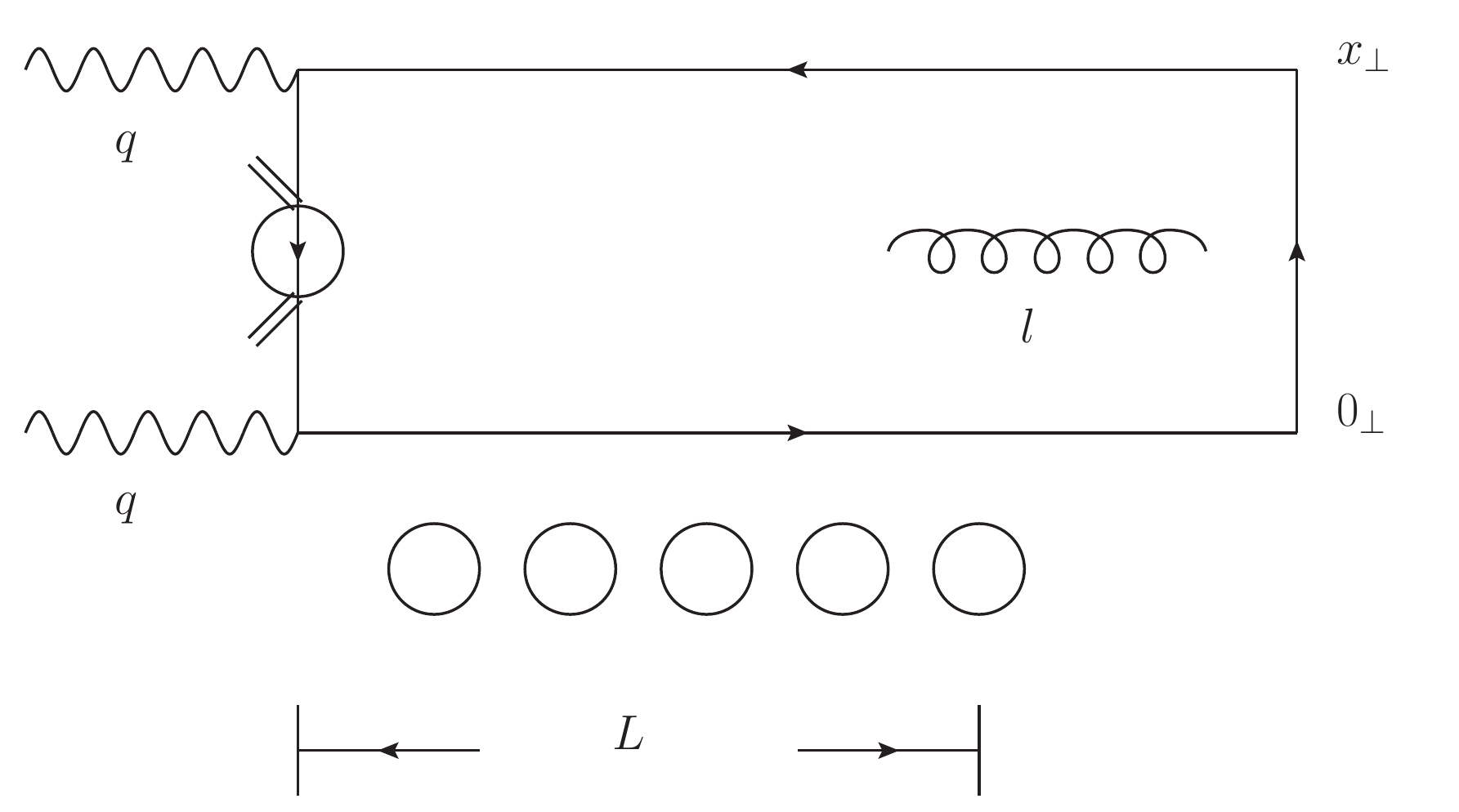}
\end{center}
\caption[*]{Forward jet production in DIS in dipole model.}
\label{f2}
\end{figure}

In order to evaluate the Sudakov term, and later the $\mathcal{E}_\textrm{MIR}$ term, it is convenient to bring the complex conjugate amplitudes in Fig.~\ref{f1} into the amplitude and view the process as in Fig.~\ref{f2}\cite{Kovchegov:2001sc, Mueller:2012bn}. In Fig.~\ref{f2} we have taken the virtual photon to interact on the front face of the nucleus so that the quark goes through a length $L$ of nuclear matter. We have also added a gauge link at $t=\infty$ to make the process manifestly gauge invariant, and we have indicated a gluon line $l$ which is emitted, and absorbed by the $0_\perp$ and $x_\perp$ quark and antiquark lines. (Emission and reabsorption of $l$ off $0_\perp$ corresponds to a virtual correction to the quark line in the amplitude of Fig.~\ref{f1}. Emission and reabsorption off $x_\perp$ corresponds to a virtual correction to the quark line in the complex conjugate amplitude of Fig.~\ref{f2} while emission off $0_\perp$ ($x_\perp$) and absorption off $x_\perp$ ($0_\perp$) corresponds to a real gluon emission correction to the graph in Fig.~\ref{f1}.)

Now the evaluation of $\mathcal{E}_{\textrm{Sud}}$ is straightforward\cite{Mueller:2012uf, Mueller:2013wwa}
\begin{equation}
\mathcal{E}_{\textrm{Sud}} =2\frac{\alpha_s C_F}{2\pi} \int_{q_+/\left[Q^2 x^2_\perp\right]}^{q_+} \frac{dl_+}{l_+} \int_{1/x_\perp^2}^{\frac{l_+}{q_+}Q^2} \frac{dl_\perp^2}{l_\perp^2} =\frac{\alpha_s C_F}{2\pi} \ln^2\left(Q^2 x^2_\perp\right). \label{sud7}
\end{equation}
The various limits to the $l_\perp^2$ and $l_+$ integration are determined as: (i) The lower limit to the $l_\perp^2$ integration comes from the fact that the softer $l_\perp$-values cancel between emissions (absorptions) off the $0_\perp$ and $x_\perp$ lines. (ii) The upper limit of the $l_\perp^2$-integration comes from the requirement that $\tau_l >\tau_q$. This is shown in some detail in appendix A. The limits on the $l_+$-integration are manifest. The logarithmic contribution given in (\ref{sud7}) comes completely from the virtual contributions as described above. The real emissions serve only to cancel the virtual emissions in the $l_\perp^2 x_\perp^2 \ll 1$ region. 

The lifetime, $\tau_l =\frac{2l_+}{l_\perp^2}$, can be either less than $L$ or greater than $L$ in (\ref{sud7}) so that the gluon, $l$, will sometimes exist within the medium. However, the gluon is too close to either the quark $0_\perp$ or antiquark ($x_\perp$) for the interactions with the medium to distinguish, say, the quark-$l$ system from the quark so that medium interactions with the gluon cancel out leaving the Sudakov term medium independent. 

It is interesting to note that the Sudakov effects occur when a dipole is created in a medium, as given by (\ref{spectrum}) and illustrated in Fig.~\ref{f2}, however there are no Sudakov effects in dipole nucleus scattering where the $t<0$ and $t>0$ regions occur in a symmetric way and there is no hard reaction to stimulate radiation. 

If $Q$ is very large then the typical values of $k_\perp$ for which $\frac{dN}{d^2 b d^2 k_\perp}$ is large will be determined by $\mathcal{E}_{\textrm{Sud}}$ given in (\ref{sud7}) and used in (\ref{spectrum}) rather than by $\hat q$ or $\hat{q}_t$\cite{Mueller:2016gko}. This is the situation for jet azimuthal angle distributions measured in ion-ion collisions at the LHC where Sudakov effects overwhelm $\hat q$ effects
\cite{Mueller:2016gko}. The interplay of Sudakov and $\hat q$ effects in (\ref{spectrum}) is an essential factor for dijet production in 
heavy ion collisions. Theoretically, in the case 
that Sudakov effects are the dominant broadening effects, 
the radiative corrections to $\hat q$ leading to $\hat{q}_t$ changes 
from the standard calculations of Refs.~\cite{Liou:2013qya,Wu:2011kc}, which will be discussed in the 
following subsection.

\subsection{Radiative corrections to $\hat q$.}
\label{rcq}
\begin{figure}[tbp]
\begin{center}
\includegraphics[width=10cm]{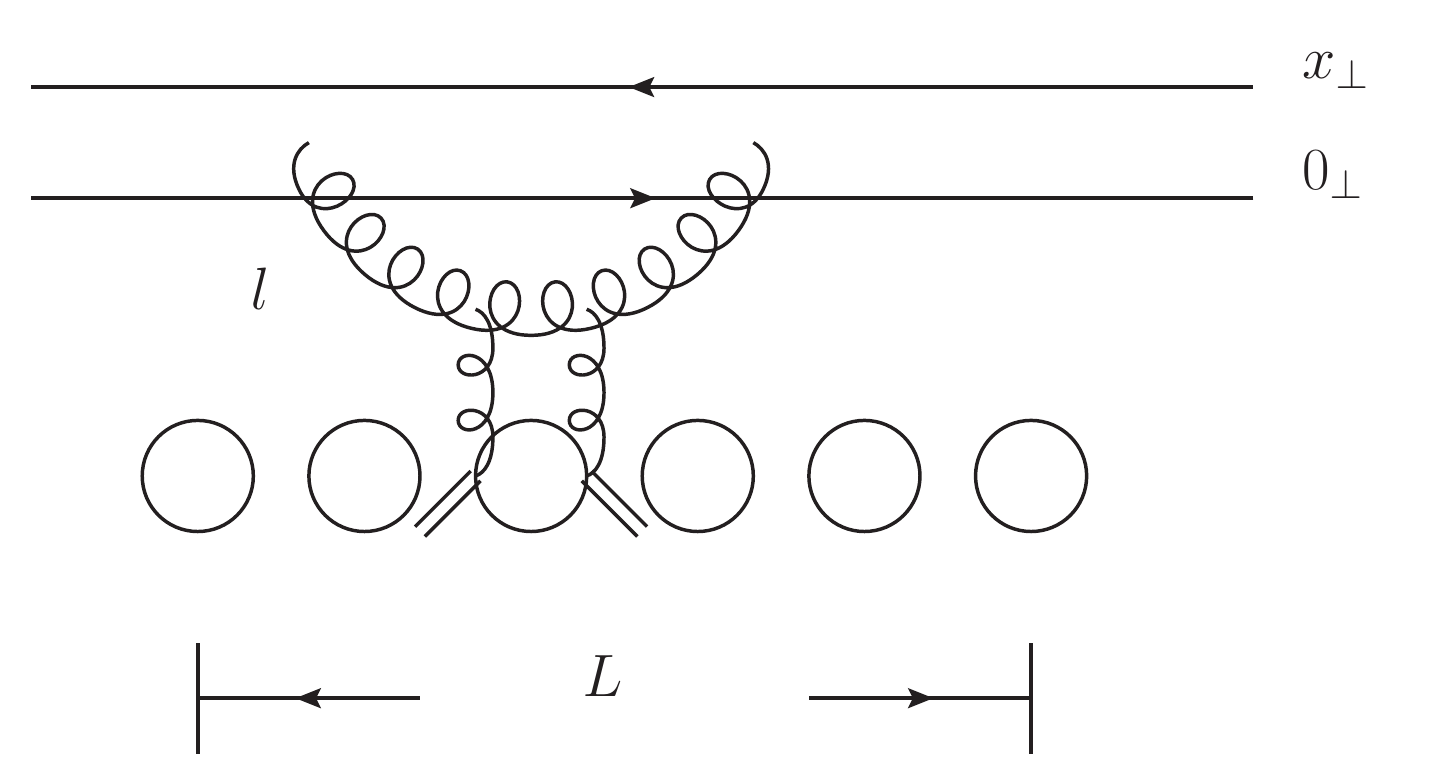}
\end{center}
\caption[*]{Radiative correction to dipole-nucleus scattering in dipole model.}
\label{f3}
\end{figure}

\begin{figure}[tbp]
\begin{center}
\includegraphics[width=10cm]{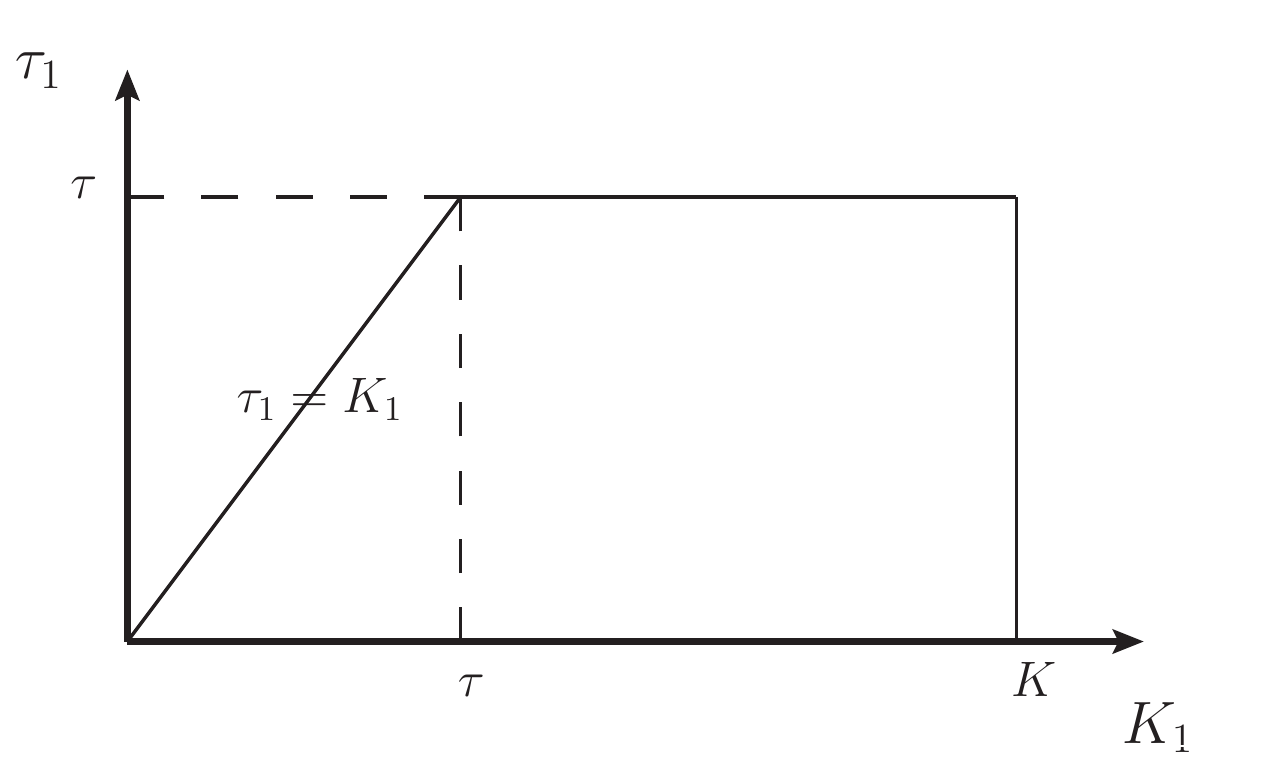}
\includegraphics[width=6.5cm]{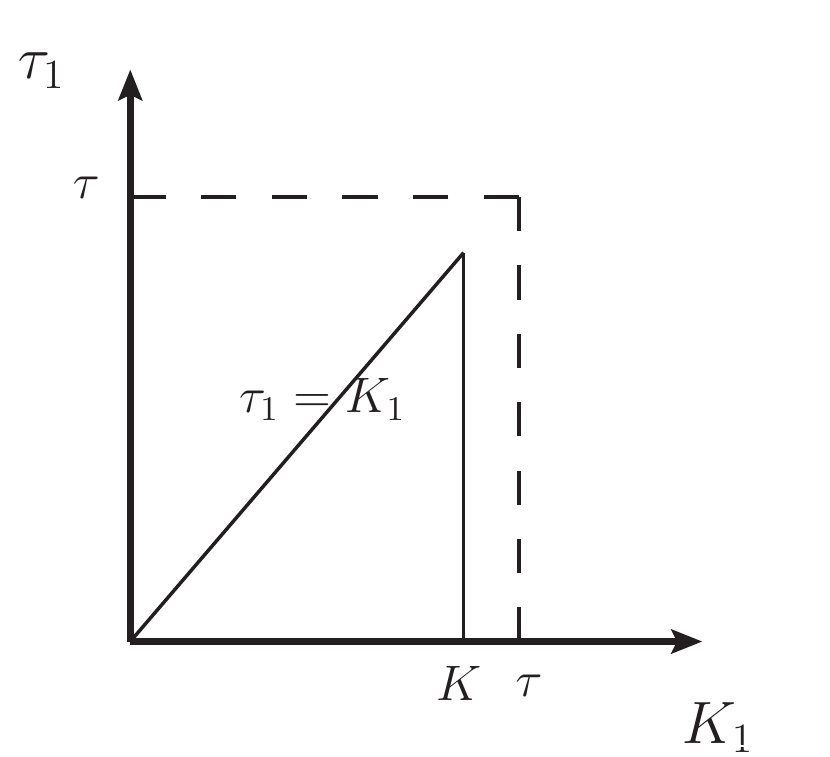}
\end{center}
\caption[*]{Domains of integration respectively for $K>\tau$ (left panel) and for $0<K\leq\tau$ (right panel).}
\label{f4}
\end{figure}

In the previous evaluation of the radiative corrections (double logarithmic) to $\hat q$~\cite{Liou:2013qya, Wu:2011kc, Blaizot:2014bha, Iancu:2014kga}, one considers gluon emission from a dipole, similar to that in Fig.~\ref{f2}. However, in this case, the gluon interacts with the medium making the effect medium dependent, as a correction to $\hat q$. The effective value of $x_\perp^2$ of the dipole is $x_\perp^2 \sim 1/(\hat q L)=1/Q_s^2$ when transverse momentum broadening is $\hat q$ dominated. If, however, the broadening is Sudakov dominated the value of $x_\perp^2$ will change and a new evaluation is necessary. At lowest order the radiative correction to $\hat q$ is illustrated in Fig.~\ref{f3} and given by 
\begin{equation}
\hat{q}_t =\hat q \left(1+ \frac{\alpha_s N_c}{\pi} \int \frac{dl_\perp^2}{l_\perp^2} \int \frac{dl_+}{l_+}\right), \label{qt}
\end{equation}
where the limits of integration have yet to be set. $\hat q$, as earlier, is the quark transport coefficient and we work in the fixed coupling approximation. The limits of integration in (\ref{qt}) are set by the following constraints:
\begin{eqnarray}
 \frac{2l_+}{l_\perp^2}&<&L \label{e9} \ ,\\
 \frac{2l_+}{l_\perp^2} &<&\frac{l_\perp^2}{\hat q}\label{e10} \ ,\\
 \frac{2l_+}{l_\perp^2} &>&r_0 \label{e11}\ ,\\
 l_\perp^2 &<&\frac{1}{x_\perp^2} \label{e12} \ ,\\
  l_+ &<& q_+  \label{e13}\ .
\end{eqnarray}
The physics meanings of the above constraints are as follows:
(\ref{e9}) is the constraint that the gluon, $l$, be within the medium; (\ref{e10}) is a single scattering requirement, necessary to get a double logarithm; (\ref{e11}) requires that the fluctuation live longer than the proton size, $r_0$; (\ref{e12}) requires that the gluon transverse distance from the dipole is greater than the dipole size, which is necessary for a double logarithm to emerge. In particular, (\ref{e10}) is a stronger requirement than (\ref{e9}) when $l_\perp^2 <\hat q L$, while (\ref{e9}) is the stronger requirement when $\hat q L < l_\perp^2 <1/x_\perp^2$. Much of what follows can also be found in \cite{Iancu:2014sha}. We include this simplified discussion for completeness.

Let us start with $1/x_\perp^2 > \hat{q}L$. Writing (\ref{qt}) more completely and using the constraints of (\ref{e9})-(\ref{e13}), we arrive at,
\begin{equation}
\hat{q}_t -\hat q = \bar{\alpha}_s \hat q \left[\int_{\hat q r_0}^{\hat q L} \frac{d l_\perp^2}{l_\perp^2}  \int_{l_\perp^2 r_0}^{(l_\perp^2)^2/\hat{q}}\frac{d l_+}{l_+}+\int_{\hat q L}^{1/x_\perp^2} \frac{d l_\perp^2}{l_\perp^2}  \int^{l_\perp^2 L}_{l_\perp^2 r_0}\frac{d l_+}{l_+}\right], \label{e14}
\end{equation}
or 
\begin{equation}
\hat{q}_t -\hat q = \bar{\alpha}_s \hat q  \ln \frac{L}{r_0} \left(\frac{1}{2} \ln \frac{L}{r_0}+\ln\frac{1}{\hat{q}L x_\perp^2}\right),
\end{equation}
where $\bar{\alpha}_s \equiv \alpha_s N_c/\pi$.
In order to sum the whole series of double logs it is convenient to introduce the following logarithmic variables
\begin{eqnarray}
K&=& \ln \frac{1}{\hat q r_0 x_\perp^2}, \quad K_1=\ln \frac{l_\perp^2}{\hat q r_0}, \\
\tau &=& \ln \frac{L}{r_0}, \quad \tau_1=\ln \frac{l_+}{l_\perp^2 r_0}\ .
\end{eqnarray}
With these notations, Eq.~(\ref{e14}) takes the form
\begin{equation}
\hat{q}_t -\hat q = \bar{\alpha}_s \hat q \int_{0}^{\tau} d\tau_1  \int_{\tau_1}^{K} dK_1 = \bar{\alpha}_s \hat q \left[K\tau -\frac{1}{2} \tau^2 \right]. \label{e18}
\end{equation}
The domain of integration for $K_1$ , $\tau_1$ in (\ref{e18}) is shown in the left panel of Fig.~\ref{f4}. The boundary $\frac{2l_+}{l_\perp^2} =\frac{l_\perp^2}{\hat q}$ given in (\ref{e10}) becomes the boundary $\tau_1 =K_1$ in Fig.~\ref{f4}. It is now straightforward to sum the complete double logarithmic series as 
\begin{equation}
\hat{q}_t =\hat q \sum_{n=0}^{\infty} \Delta_n, \label{sum}
\end{equation}
where 
\begin{equation}
\Delta_n =\Pi_{i=1}^n \bar{\alpha}_s \int_{0}^{\tau_{i+1}} d\tau_i  \int_{\tau_i}^{K_{i+1}} dK_i \, \label{e20}
\end{equation}
with $\tau_{n+1}=\tau$ and $K_{n+1}=K$ in (\ref{e20}). Therefore, we find that $\Delta_n$ obeys 
the following equation,
\begin{equation}
\frac{\partial}{\partial \tau}\frac{\partial}{\partial K} \Delta_n (\tau, K) = \bar{\alpha}_s \Delta_{n-1} (\tau, K), \label{e21}
\end{equation}
which, with (\ref{e18}), gives
\begin{equation}
\Delta_n = \frac{\bar{\alpha}_s^n K^{n-1}\tau^n \left[(n+1)K-n\tau\right]}{n! (n+1)!}. \label{e22}
\end{equation}
Using (\ref{e22}), the sum in (\ref{sum}) can be derived~\cite{Iancu:2014sha} 
\begin{equation}\label{qtlargeK}
\hat{q}_t =\hat{q} \left[ \frac{1}{\sqrt{\bar{\alpha}_s K\tau }} I_1 (2 \sqrt{\bar{\alpha}_s K\tau }) +\left(1- \frac{\tau}{K}\right) I_2 (2 \sqrt{\bar{\alpha}_s K\tau })  \right].
\end{equation}

In the case of $\hat{q}r_0\leq1/x_\perp^2 \leq \hat{q}L$, i.e., $0\leq K \leq \tau$, the domain of integration for $K_1$ is shown in the right panel of Fig. \ref{f4}. With that, we find that Eq.~(\ref{qt}) can be written as
\begin{equation}
\hat{q}_t -\hat q = \bar{\alpha}_s \hat q \int_{0}^{K} d\tau_1  \int_{\tau_1}^{K} dK_1 = \frac{1}{2}\bar{\alpha}_s \hat q K^2,
\end{equation}
which is simply given by (\ref{e18}) with $\tau$ being replaced by $K$. Similarly, it is easy to see, for this case, that
\begin{equation}
\Delta_n =\frac{\bar{\alpha}_s^n K^{2n}}{n! (n+1)!}, \quad
\textrm{and}  \quad
\hat{q}_t =\hat{q} \frac{1}{\sqrt{\bar{\alpha}_s }K} I_1 (2 \sqrt{\bar{\alpha}_s }K).
\end{equation}
In summary, we have the following results for different $K$ values,
\begin{subequations}
\label{equations}
\begin{align}
  \label{eqa}
 \hat{q}_t  &=\hat{q}  &\text{if $K<0$,}\vspace{2mm}\\
  \label{eqb}
   \hat{q}_t  &=\hat{q}   \frac{1}{\sqrt{\bar{\alpha}_s }K} I_1 (2 \sqrt{\bar{\alpha}_s }K)&\text{if $0\leq K\leq\tau$,}\vspace{2mm}\\
   \label{eqc}
   \hat{q}_t  &= \hat{q} \left[ \frac{1}{\sqrt{\bar{\alpha}_s K\tau }} I_1 (2 \sqrt{\bar{\alpha}_s K\tau }) +\left(1- \frac{\tau}{K}\right) I_2 (2 \sqrt{\bar{\alpha}_s K\tau })  \right]& \text{if $K>\tau$.}
\end{align}
\end{subequations}
The above results, together with (\ref{sud7}),
give the complete evaluation of $\mathcal{E}$ in (\ref{spectrum}) via (\ref{tot}). (When used in (\ref{spectrum}) the $L$ in $\tau$ should be changed to $z$ the effective path length for the integrand of (\ref{spectrum}).)

The spectrum $\frac{dN}{d^2 b d^2 k_\perp}$ is then given by (\ref{spectrum}) or (\ref{wwquark}) where all the ingredients for evaluating (\ref{spectrum}) or (\ref{wwquark}) are given by (\ref{sud7}) and (\ref{equations}). It is straightforward to include running coupling effects and higher order corrections to the Sudakov term in (\ref{sud7}). (See (\ref{running43}) for running coupling corrections.) However, it is not clear at present how to include running coupling corrections to $\hat{q}_t$ in a resummed way. See Ref.~\cite{Iancu:2014sha} for the state of the art. 

More interestingly, following (\ref{e21}) and summing over all $n$, one can in fact write down a double differential evolution equation for $\hat{q}_t $ as follows
\begin{equation}
 \frac{\partial}{\partial \tau}\frac{\partial}{\partial K} \hat{q}_t = \bar{\alpha}_s \hat{q}_t, \label{qe}
\end{equation}
which is equivalent to the DGLAP evolution equation for the gluon distribution in the double logarithmic limit. As shown in Ref.~\cite{Gorshkov:1966ht, Kovchegov:2012mbw}, the solution of (\ref{qe}) can be written in terms of superpositions of modified Bessel functions $I_{\nu} (x)$ with coefficients determined by boundary conditions, since $\left(\frac{\tau}{\kappa}\right)^{\frac{\nu}{2}}I_\nu(2\sqrt{\bar\alpha_s\tau\kappa})$ for arbitrary $\nu$ is a solution to (\ref{qe}). For example, given the boundary conditions $\hat{q}_t|_{K=0}=\hat{q}_t|_{\tau=0}=\hat{q}$, one can find $ \hat{q}_t  =  \hat{q} I_0(2 \sqrt{\bar{\alpha}_s K\tau })$ which is equivalent to the usual DGLAP double logarithmic solution for gluon distributions.\footnote{This is natural since it is known that $\hat{q}$ is proportional to target gluon distributions by definition.} Furthermore, it is straightforward to check that (\ref{qtlargeK}) or (\ref{eqc}) is the solution to (\ref{qe}) given boundary conditions $\hat{q}_t|_{K=0}=\hat{q} (1-\frac{\bar{\alpha}_s}{2} \tau^2 )$ and $\hat{q}_t |_{\tau=0}=\hat{q}$. This indicates that the evolution equation of $\hat{q}_t$ in the double logarithmic limit is also given by (\ref{qe}) with particular boundary conditions which reflect the information of the target medium such as length $L$ and multiple scatterings. Therefore, it seems that we can obtain the full results in (\ref{equations}) by continuing the solution (\ref{eqc}) to (\ref{eqb}) at $K=\tau$ then to (\ref{eqa}) at $K=0$.

\section{Small-$x$ forward jet production in DIS}
\label{smc}

\begin{figure}[tbp]
\begin{center}
\includegraphics[width=10cm]{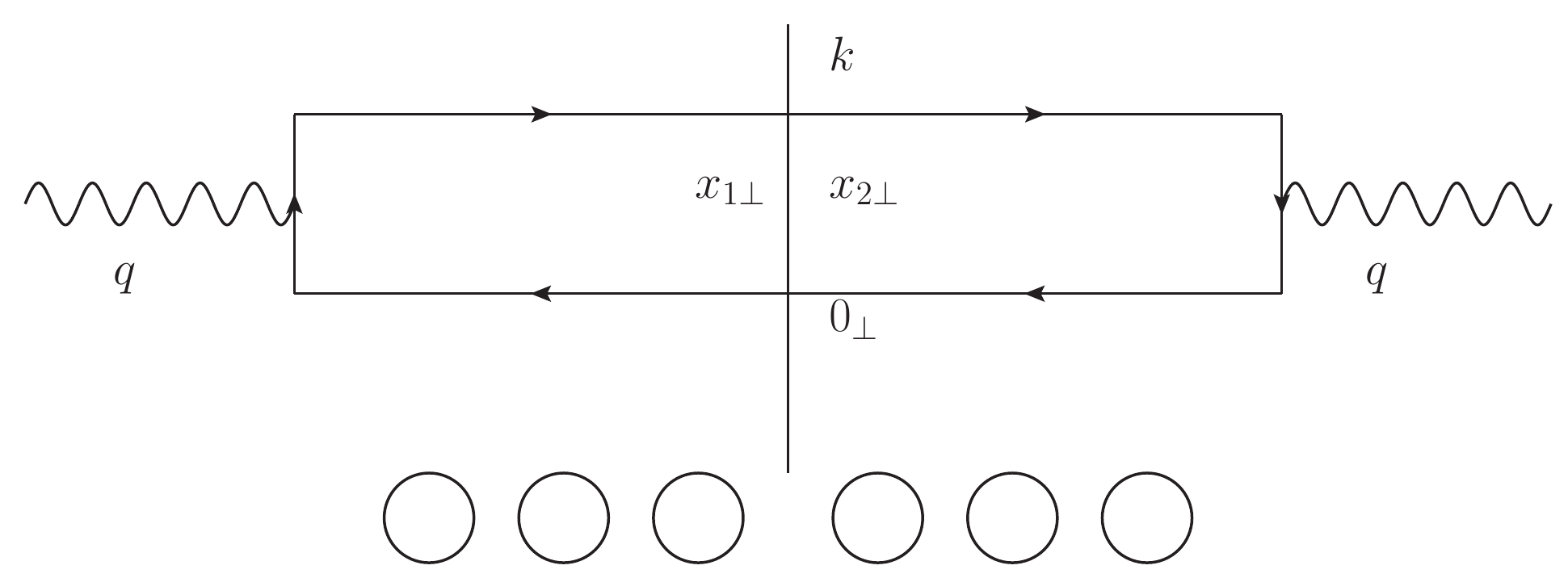}
\end{center}
\caption[*]{Radiative correction to dipole-nucleus scattering in dipole model in the small-$x$ limit.}
\label{f5}
\end{figure}

\begin{figure}[tbp]
\begin{center}
\includegraphics[width=10cm]{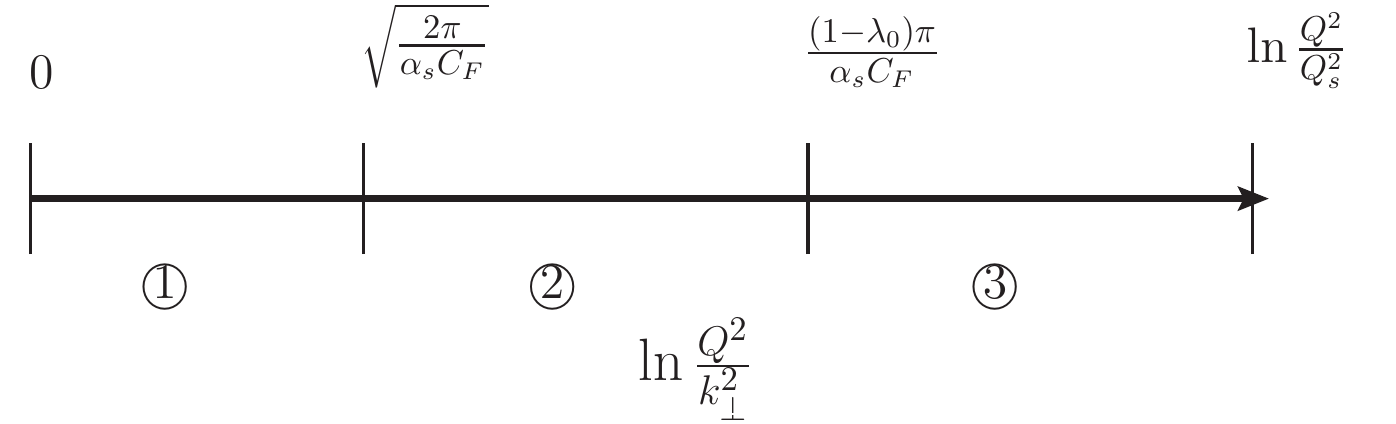}
\end{center}
\caption[*]{Three regions of the transverse momentum spectrum as a function of $\ln\frac{Q^2}{k_\perp^2}$ in case $\tau_q\gg L$.}
\label{f6}
\end{figure}

We now turn to the limit opposite to that of the large medium considered in Sec. \ref{lm}, namely the case where $\tau_q \gg  L$. Here the process has the virtual photon splitting into a quark-antiquark dipole which then further evolves before passing over the nucleus. The process is illustrated in Fig.~5 where evolution of the ($x_{\perp 1}, 0_\perp$) and ($x_{\perp 2}, 0_\perp$) dipoles in the amplitude and complex conjugate amplitude are not explicitly shown, nor are the interactions with the target nucleus shown. 

\subsection{The forward jet spectrum; fixed coupling analysis}
\label{fixed}
The forward quark (or antiquark) jet spectrum coming from the scattering of a transverse virtual photon is usually written as \cite{Mueller:1999wm}
\begin{eqnarray}
\frac{dN}{d^2b d^2 k_\perp} &=&\sum_f e^2_f \frac{Q^2 N_c}{32\pi^6} \int d^2 x_{1\perp }d^2 x_{2\perp }e^{-ik_\perp \cdot (x_{1\perp}-x_{2\perp})} \int_0^1 dz \left[z^2+(1-z)^2\right]  \times  \notag \\
&&\nabla_{x_{1\perp}} K_0 \left[\sqrt{Q^2 x_{1\perp}^2 z(1-z)}\right]\cdot \nabla_{x_{2\perp}} K_0 \left[\sqrt{Q^2 x_{2\perp}^2 z(1-z)}\right] \left[1+S(x_{1\perp} -x_{2\perp})-S(x_{1\perp})-S(x_{2\perp})\right]. \label{smallxe24}
\end{eqnarray}
The different factors in (\ref{smallxe24}) are straightforward to understand: the $\nabla K_0$ factors are the quark-antiquark wavefunctions of the virtual transverse photons in the amplitudes and complex conjugate amplitude; $\left[z^2+(1-z)^2\right]$ is the splitting function of the photon. For the various $S$-matrices the combination in (\ref{smallxe24}) guarantees that there be at least one interaction in the amplitude and in the complex conjugate amplitude. The normalization is 
\begin{equation}
\int d^2b d^2k_\perp \frac{dN}{d^2b d^2 k_\perp} =\sum_f e^2_f \left[ xq_A^f (x, Q^2)+x\bar{q}_A^f (x, Q^2)\right]. \label{norm}
\end{equation}

However, (\ref{smallxe24}) is missing a Sudakov factor. One often says that DIS scattering is given in terms of a dipole scattering amplitude times the virtual photons' quark-antiquark wave functions. That is true of (\ref{norm}) with $\frac{dN}{d^2b d^2 k_\perp}$ given by (\ref{smallxe24}). However if a jet is measured rather than integrated over, as in (\ref{norm}), a Sudakov factor \cite{Mueller:2013wwa}
\begin{equation}
\textrm{Sudakov} = e^{-\frac{\alpha_s C_F}{2\pi} \ln^2 \left[Q^2(x_{1\perp}-x_{2\perp})^2+1\right]} \label{sud26}
\end{equation} 
should be inserted in the integrand in (\ref{smallxe24}). (The $1$ in $Q^2(x_{1\perp}-x_{2\perp})^2+1$ is included to make the $x_{1\perp} \to x_{2\perp}$ limit, as occurs in (\ref{norm}), non-singular.) Now inserting (\ref{sud26}) into the integrand of (\ref{smallxe24}) and changing the variables of integration, we will                  get 
\begin{eqnarray}
\frac{dN}{d^2b d^2 k_\perp} &=&\sum_f e^2_f \frac{Q^2 N_c}{32\pi^6} \int d^2 x_{1\perp }d^2 x_{\perp } \int_0^1 dz  e^{-ik_\perp \cdot x_{\perp}}\left[z^2+(1-z)^2\right]  \nabla_{x_{1\perp}} K_0 \left[\sqrt{Q^2 x_{1\perp}^2 z(1-z)}\right]  \notag \\
&&\cdot \nabla_{x_{1\perp}} K_0 \left[\sqrt{Q^2 (x_{1\perp}-x_{\perp})^2 z(1-z)}\right] e^{-\frac{\alpha_s C_F}{2\pi} \ln^2 (Q^2 x_{\perp}^2)}\left[1+S(x_{\perp})-2S(x_{1\perp})\right]. \label{sudx27}
\end{eqnarray}
While it appears difficult to do the $x_{1\perp}$-integration in (\ref{sudx27}) exactly, it is clear that $|x_{1\perp}|\sim |x_\perp|$ dominates the leading power contribution of the integral and that $z\sim \frac{1}{Q^2 x_{\perp}^2}$ or $1-z \sim \frac{1}{Q^2 x_{\perp}^2}$ so that 
\begin{equation}
\frac{dN}{d^2b d^2 k_\perp} =\mathcal{C}\int \frac{d^2 x_\perp}{\pi x_\perp^2} e^{-ik_\perp\cdot x_\perp} e^{-\frac{\alpha_s C_F}{2\pi} \ln^2 (Q^2 x_{\perp}^2)} T(x_\perp, Y) \label{e28}
\end{equation}
where $1-S=T$ and $Y=\ln\frac{1}{x_{\textrm{Bj}}}$. $\mathcal{C}$ is a constant in the sense that it has no $k_\perp$-dependence but it will depend on the form of $T$, that is in the scaling region it will depend on the anomalous dimension giving the scaling behavior. As an illustration, let us suppose that the energy is high enough that $k_\perp$ can be taken in the scaling region \cite{Kwiecinski:2002ep, Iancu:2002tr}
\begin{equation}
T(x_\perp )=\left[ Q_s^2 x_\perp^2\right]^{1-\lambda_0}.
\end{equation}
It is straightforward to get 
\begin{equation}
\frac{dN}{d^2b d^2 k_\perp} =2\mathcal{C}\int_0^{\infty} \frac{d x_\perp}{ x_\perp} \left(Q_s^2 x_\perp^2\right)^{1-\lambda_0} J_0(k_\perp x_\perp) e^{-\frac{\alpha_s C_F}{2\pi} \ln^2 (Q^2 x_{\perp}^2)}
\end{equation}
or
\begin{equation}
\frac{dN}{d^2b d^2 k_\perp} =2\mathcal{C} \left(\frac{Q_s^2}{Q^2}\right)^{1-\lambda_0}\int_0^{\infty} \frac{d x_\perp}{ x_\perp} J_0(k_\perp x_\perp) e^{-\frac{\alpha_s C_F}{2\pi} \ln^2 (Q^2 x_{\perp}^2)+(1-\lambda_0) \ln (Q^2 x_{\perp}^2)}.
\end{equation}
Changing variables to $z=\ln (Q^2 x_{\perp}^2)$ one gets 
\begin{equation}
\frac{dN}{d^2b d^2 k_\perp} =\mathcal{C}  \left(\frac{Q_s^2}{Q^2}\right)^{1-\lambda_0}\int_{-\infty}^{+\infty} dz J_0\left(\frac{k_\perp}{Q} e^{z/2} \right)e^{-\frac{\alpha_s C_F}{2\pi} z^2 +(1-\lambda_0)z }. \label{e32}
\end{equation}
Although the $z$ integration has be written as going from $-\infty$ to $\infty$ in (\ref{e32}), the effective range can be taken as $z< \ln \frac{Q^2}{k_\perp^2}$
\begin{equation}
\frac{dN}{d^2b d^2 k_\perp} =\mathcal{C}  \left(\frac{Q_s^2}{Q^2}\right)^{1-\lambda_0}\int_{-\infty}^{\ln \frac{Q^2}{k_\perp^2}} dz e^{-\frac{\alpha_s C_F}{2\pi} z^2 +(1-\lambda_0)z }. \label{e33}
\end{equation}
It is now straightforward to write (\ref{e33}) in terms of the error function $Erf(x)$ by rewriting (\ref{e33}) as 
 \begin{equation}
\frac{dN}{d^2b d^2 k_\perp} =\mathcal{C}  \left(\frac{Q_s^2}{Q^2}\right)^{1-\lambda_0}e^{\frac{(1-\lambda_0)^2 \pi}{2\alpha_s C_F}} \sqrt{\frac{2\pi}{\alpha_s C_F}} \int_{-\infty}^{\sqrt{\frac{\alpha_s C_F}{2\pi}} \left[\ln \frac{Q^2}{k_\perp^2}-\frac{(1-\lambda_0) \pi}{\alpha_s C_F}\right]} dw e^{-w^2 }, \label{e34}
\end{equation}
where we have introduced 
\begin{equation}
w=\sqrt{\frac{\alpha_s C_F}{2\pi}} \left[z-\frac{(1-\lambda_0) \pi}{\alpha_s C_F}\right].
\end{equation}
Using 
\begin{equation}
Erf (x) =\int_0^x dt e^{-t^2},
\end{equation}
one has 
 \begin{equation}
\frac{dN}{d^2b d^2 k_\perp} =\mathcal{C}  \left(\frac{Q_s^2}{Q^2}\right)^{1-\lambda_0}e^{\frac{(1-\lambda_0)^2 \pi}{2\alpha_s C_F}} \sqrt{\frac{2\pi}{\alpha_s C_F}} \left\{ Erf \left[\sqrt{\frac{\alpha_s C_F}{2\pi}} \left(\ln \frac{Q^2}{k_\perp^2}-\frac{(1-\lambda_0) \pi}{\alpha_s C_F}\right)\right] +\frac{\sqrt{\pi}}{2} \right\}. \label{e37}
\end{equation}
One can write approximate results for the three regions shown in Fig.~\ref{f6} as follows:
\begin{eqnarray}
\left.\frac{dN}{d^2b d^2 k_\perp}\right|_{\circled{1}}&=&\frac{\mathcal{C}}{1-\lambda_0}\left(\frac{Q_s^2}{k_\perp^2}\right)^{1-\lambda_0}, \label{e39} \\
\left.\frac{dN}{d^2b d^2 k_\perp}\right|_{\circled{2}}&=&\frac{\mathcal{C}}{1-\lambda_0}\left(\frac{Q_s^2}{k_\perp^2}\right)^{1-\lambda_0} e^{-\frac{\alpha_s C_F}{2\pi} \ln^2 \frac{Q^2}{k_\perp^2}}, \label{e40}\\
\left.\frac{dN}{d^2b d^2 k_\perp}\right|_{\circled{3}}&=&\pi \mathcal{C} \sqrt{\frac{2}{\alpha_s C_F}} e^{\frac{(1-\lambda_0)^2 \pi}{2\alpha_s C_F}}  \left(\frac{Q_s^2}{Q^2}\right)^{1-\lambda_0}. \label{e41}
\end{eqnarray}
Equations (\ref{e39}-\ref{e41}) are approximate equations, accurate away from the boundaries of their respective regions. To get an accurate evaluation at the boundary of regions $\circled{2}$ and $\circled{3}$ one must use (\ref{e37}). Equations (\ref{e39}) and (\ref{e40}) have a smooth transition between regions $\circled{1}$ and $\circled{2}$ so that (\ref{e40}) can be used in both regions. Also, if $\ln \frac{Q^2}{Q_s^2} <\frac{(1-\lambda_0) \pi}{\alpha_s C_F}$ region $\circled{3}$  does not exist and (\ref{e40}) again becomes the relevant formula. 

In region $\circled{1}$ Sudakov effects are very small and the "normal" scaling result holds as indicated in (\ref{e39}). When $k_\perp^2$ decreases, one moves to region $\circled{2}$ where Sudakov effects appear in a very simple way, modifying the geometric scaling formula. Finally, if $\ln \frac{Q^2}{Q_s^2}$ is large enough, when $\ln \frac{Q^2}{k_\perp^2}$ gets larger than $\frac{(1-\lambda_0) \pi}{\alpha_s C_F}$ the $k_\perp$ of the jet does not come mainly from small-$x$ evolution and all $k_\perp$-dependence has disappeared due to the randomizing effects of Sudakov radiation.
 
\subsection{The forward jet spectrum; running coupling}
\label{rcs}

Now we shall repeat the discussion given in Sec.~\ref{fixed} but using a running QCD coupling and without the assumption that $Q^2$ is in the scaling region of the small-$x$ evolution. Not too much changes from our earlier analysis and much of the discussion of Sec.~\ref{fixed} can be directly taken over to the running coupling case. The main change is the modification of (\ref{sud26}). Now Sudakov effects take the form
\begin{equation}
\ln \textrm{Sud.} =- \frac{C_F}{\pi} \int_{1/x_\perp^2}^{Q^2} \frac{dl_\perp^2}{l_\perp^2} \alpha_s (l_\perp^2) \int^{q_+}_{q_+\frac{l_\perp^2}{Q^2}} \frac{dl_+}{l_+}.
\end{equation}
Using $\alpha_s =\frac{1}{b\ln l_\perp^2/\Lambda^2}$, one finds 
\begin{equation}
\ln \textrm{Sud.}=-\frac{C_F}{\pi b}\left[ \ln \frac{Q^2}{\Lambda^2} \ln\frac{\ln \frac{Q^2}{\Lambda^2} }{\ln  \frac{1}{x_\perp^2\Lambda^2}}-\ln Q^2 x_\perp^2\right] \label{running43}
\end{equation}
which now replaces (\ref{sud26}). Except for the Sudakov factor, (\ref{sudx27}) and (\ref{e28}) still hold. It is straightforward to write (\ref{running43}) in terms of $z=\ln Q^2 x_\perp^2 $ as 
\begin{equation}
\ln \textrm{Sud.}=\frac{C_F}{\pi b}\left[ \ln \frac{Q^2}{\Lambda^2} \ln\left(1-\frac{z}{ \ln\frac{Q^2}{\Lambda^2}}\right)+z\right] \label{e44}
\end{equation}
and, expanding the logarithm and using $\alpha_s (Q)=\frac{1}{b\ln Q^2/\Lambda^2}$, to get 
\begin{equation}
\ln \textrm{Sud.}=-\frac{C_F}{\pi b^2 \alpha_s (Q)} \sum_{n=2}^{\infty}\frac{\left(zb\alpha_s (Q)\right)^n}{n} \label{sudex}
\end{equation}
or 
\begin{equation}
\ln \textrm{Sud.}=-\frac{\alpha_s (Q) C_F}{2\pi }z^2 +\cdots \label{sudexpand}
\end{equation}
in case $zb\alpha_s (Q)$ is small. 

The regions $\circled{1}, \circled{2}, \circled{3}$ in Fig. ~6 are essentially unchanged except for the value of $\ln \frac{Q^2}{k_\perp^2}$ separating regions $\circled{2}$ from $\circled{3}$. Let us begin near $\ln \frac{Q^2}{k_\perp^2}=0$, the left-most region in Fig.~6 where (\ref{e33}) now becomes 
\begin{equation}
\frac{dN}{d^2b d^2 k_\perp} =\mathcal{C}  \int_{0}^{\ln \frac{Q^2}{k_\perp^2}} dz e^{-\frac{\alpha_s C_F}{2\pi} z^2 +\ln T(z, Q, Y) } \label{e47}
\end{equation}
where $Y$, in $T(z, Q, Y)$, is $Y=\ln\frac{1}{x_{\textrm{Bj}}}$. When $\ln \frac{Q^2}{k_\perp^2}$ is not too large $z\alpha_s (Q) \ll 1 $ and we have used (\ref{sudexpand}) as the Sudakov factor. So long as $\frac{\alpha_s C_F}{2\pi} \ln^2 \frac{Q^2}{k_\perp^2} <1 $, region $\circled{1}$, the Sudakov factor in (\ref{e47}) may be dropped and 
\begin{equation}
\frac{dN}{d^2b d^2 k_\perp} =\mathcal{C}  \int_{0}^{\ln \frac{Q^2}{k_\perp^2}} dz \, T(z, Q, Y).  \label{e48}
\end{equation}
If $Q$ were in the scaling region, $T(x_\perp )=\left[ Q_s^2 x_\perp^2\right]^{1-\lambda_0}$, (39) would emerge but, in any case, in region $\circled{1}$ the result for $\frac{dN}{d^2b d^2 k_\perp}$ is the usual result with Sudakov effects being negligible. 

As one moves into regions $\circled{2}$, (\ref{e47}) remains valid so long as $\ln \frac{Q^2}{k_\perp^2}$ is not too large. As in (\ref{e48}) the $z$ integration is dominated by the upper end, $z\sim \ln \frac{Q^2}{k_\perp^2}$, of the integral so that 
\begin{equation}
\frac{dN}{d^2b d^2 k_\perp} \sim e^{-\frac{\alpha_s (Q) C_F}{2\pi} \ln^2 \frac{Q^2}{k_\perp^2}} T(x_\perp =\frac{1}{k_\perp}). \label{e49}
\end{equation}
If $k_\perp$ is in the scaling region then (\ref{e40}) will emerge. Here the Sudakov correction is a simple factor times the usual result without including Sudakov effects. The transition between $\circled{2}$ and $\circled{3}$ is determined by 
\begin{equation}
\frac{d}{dz} \left( \ln \textrm{Sud.} +\ln T\right) =0.
\end{equation}
Using (\ref{e44}) this gives the equation
\begin{equation}
\frac{zb\alpha_s (Q)}{1-zb\alpha_s(Q)} =\frac{\pi b}{C_F} \frac{1}{T} \frac{\partial T}{\partial z}.\label{e51}
\end{equation}
(In case $T=(Q_s^2 x_\perp^2)^{1-\lambda_0}=\left(\frac{Q_s^2}{Q^2}\right)^{1-\lambda_0} e^{(1-\lambda_0)z}$, the scaling region, and if one used (\ref{sudexpand}) the boundary $\ln \frac{Q^2}{k_\perp^2}=\frac{(1-\lambda_0)\pi}{\alpha_s C_F}$ shown in Fig.~6 would emerge.)

Without assuming that the boundary between regions $\circled{2}$ and $\circled{3}$ is in the scaling region we can parametrize $T$ as 
\begin{equation}
T\propto (x_\perp^2)^{1-\lambda_{\textrm{eff}}} \sim e^{(1-\lambda_{\textrm{eff}})z}, \label{e52}
\end{equation}
where $\lambda_{\textrm{eff}}$ may depend on $Y$ and on $x_\perp$. Assuming the $z$ dependence of $\lambda_{\textrm{eff}}$ is small, a reasonable assumption, using (\ref{e51}) and (\ref{e52}) gives
\begin{equation}
\ln \frac{Q^2}{k_\perp^2} =\frac{(1-\lambda_{\textrm{eff}}) \pi}{C_F \alpha_s(Q) \left[1+\frac{\pi b (1-\lambda_{\textrm{eff}})}{C_F}\right]}\equiv z_0 \label{e53}
\end{equation}
as the boundary between regions $\circled{2}$ and $\circled{3}$. (The boundary between $\circled{2}$ and $\circled{3}$ as given by (\ref{e53}) is to the left of the end point, $\ln \frac{Q^2}{Q_s^2}$, in Fig.~\ref{f6} so long as $\ln \frac{Q^2}{\Lambda^2} >\left[1+\frac{\pi b (1-\lambda_{\textrm{eff})}}{C_F}\right] \ln \frac{Q_s^2}{\Lambda^2}$, which we suppose to be the case.)

In region $\circled{3}$ all $k_\perp$-dependence is lost because the Sudakov factor cuts off the $z$ integration before the upper limit, $\ln \frac{Q^2}{k_\perp^2}$, is reached. It is in this upper limit that the $k_\perp$-dependence resides. Generalizing (\ref{e47}) to read 

\begin{equation}
\frac{dN}{d^2b d^2 k_\perp} =\mathcal{C}\int_0^{\ln \frac{Q^2}{k_\perp^2}} dz e^{\ln \text{Sud} +\ln T}\label{e54}
\end{equation}
or, using (\ref{e44})
\begin{equation}
\frac{dN}{d^2b d^2 k_\perp} =\frac{\mathcal{C}}{b\alpha_s}\int_0^{1} dw e^{\frac{C_F}{\pi b^2 \alpha_s} \left[\ln(1-w)+w\right]}T(w, Q, Y) \label{e55}
\end{equation}
where $b\alpha_s z =w$. Doing the $w$-integration by integrating about the saddle point (\ref{e53}) gives
\begin{equation}
\frac{dN}{d^2b d^2 k_\perp} = \frac{\pi \mathcal{C}}{1+\frac{\pi b (1-\lambda_{\textrm{eff}})}{C_F}}\sqrt{\frac{2}{\alpha_s C_F}}e^{\frac{C_F}{\pi b^2 \alpha_s} \left[\ln(1-w_0)+w_0\right]}T(z_0, Q, Y) \label{e56}
\end{equation}
with $w_0=b\alpha_s z_0$ with $z_0$ given by (\ref{e53}). Eq.~(\ref{e41}) is recovered if one only keeps the quadratic term in $w_0$ in the exponential in (\ref{e56}) and if one takes a scaling solution for $T$.

\subsection{Medium effects}
\label{meq}

In the case of $\tau_q \ll L$, the medium effects, i.e., the multiple scattering and radiative corrections interacting with the nucleus, led to explicit nuclear medium effect summarized in (\ref{equations}) which can directly affect the spectrum and, if $Q^2$ is not too large, compete with Sudakov effects when (\ref{spectrum}) or (\ref{wwquark}) is used to evaluate the spectrum. In the small-$x$ limit such medium effects must be hidden in the $T$ in (\ref{e48}) or (\ref{e49}). The usual way to incorporate multiple scattering effects into $T$ is to use the McLerran-Venugopalan (MV) model 
\begin{equation}
T_{\textrm{MV}} (x_\perp) =1-e^{-Q^2_s(\textrm{MV})x_\perp^2/4} \label{mv}
\end{equation}
as the initial condition for the evolution of $T(x_\perp, Y)$ using the Balitsky-Kovchegov (BK) equation. The evolution is done from $Y_0$ to $Y=\ln\frac{1}{x_{Bj}}=\ln s x_\perp^2$ and where $Y_0$ is determined by requiring that the coherence of the dipole, $x_\perp$, be the nuclear length $L=2\sqrt{R^2-b^2}$ with $R$ the nuclear radium and $b$ the impact parameter. One easily determines $Y_0$ is 
\begin{equation}
Y_0 =\ln LM
\end{equation}
with $M$ the nucleon mass. 

What is missing in the above discussion is evolution in the medium. The $Q^2_s(\textrm{MV})$ in (\ref{mv}) is given by  
\begin{equation}
Q^2_s(\textrm{MV}) = \hat q L
\end{equation}
with $\hat q$ given in (\ref{qhatdef}). One can incorporate evolution in the medium, evolution below $Y_0$, simply by replacing $Q^2_s(\textrm{MV})$ in (\ref{mv}) by $Q^2_{s\, \textrm{initial}}=Q_{s \, \textrm{in}}^2$ where 
\begin{equation}
Q_{s \, \textrm{in}}^2 =\hat{q}_t L
\end{equation}
with $\hat{q}_t$ given in (\ref{equations}). Now, in contrast to $Q^2_s(\textrm{MV})$, $Q_{s \, \textrm{in}}^2 $ has a strong dependence on the dipole size and additional dependence on medium length due to double logarithmic corrections. Thus the initial condition, at $Y_0$, 
\begin{equation}
T_{\textrm{in}} (x_\perp) =1-e^{-Q^2_{s\, \textrm{in}}x_\perp^2/4} \label{in}
\end{equation}
should include evolution in the medium, at least in the fixed coupling limit. In the weak coupling limit $\bar{\alpha}_s \to 0$, $Q_{s \, \textrm{in}}^2 \to Q^2_s(\textrm{MV})$ ($\hat{q}_t \to \hat q$) and the above initial condition reduces to the MV initial condition since all the medium evolution is negligible. In general, we believe (\ref{in}) is an improved initial condition for BK-JIMWLK \cite{Balitsky:1995ub, Kovchegov:1999yj, JalilianMarian:1997gr, Iancu:2001ad} evolution compared to (\ref{mv}).

\section{Conclusion}
\label{con}

\begin{figure}[tbp]
\begin{center}
\includegraphics[width=8cm]{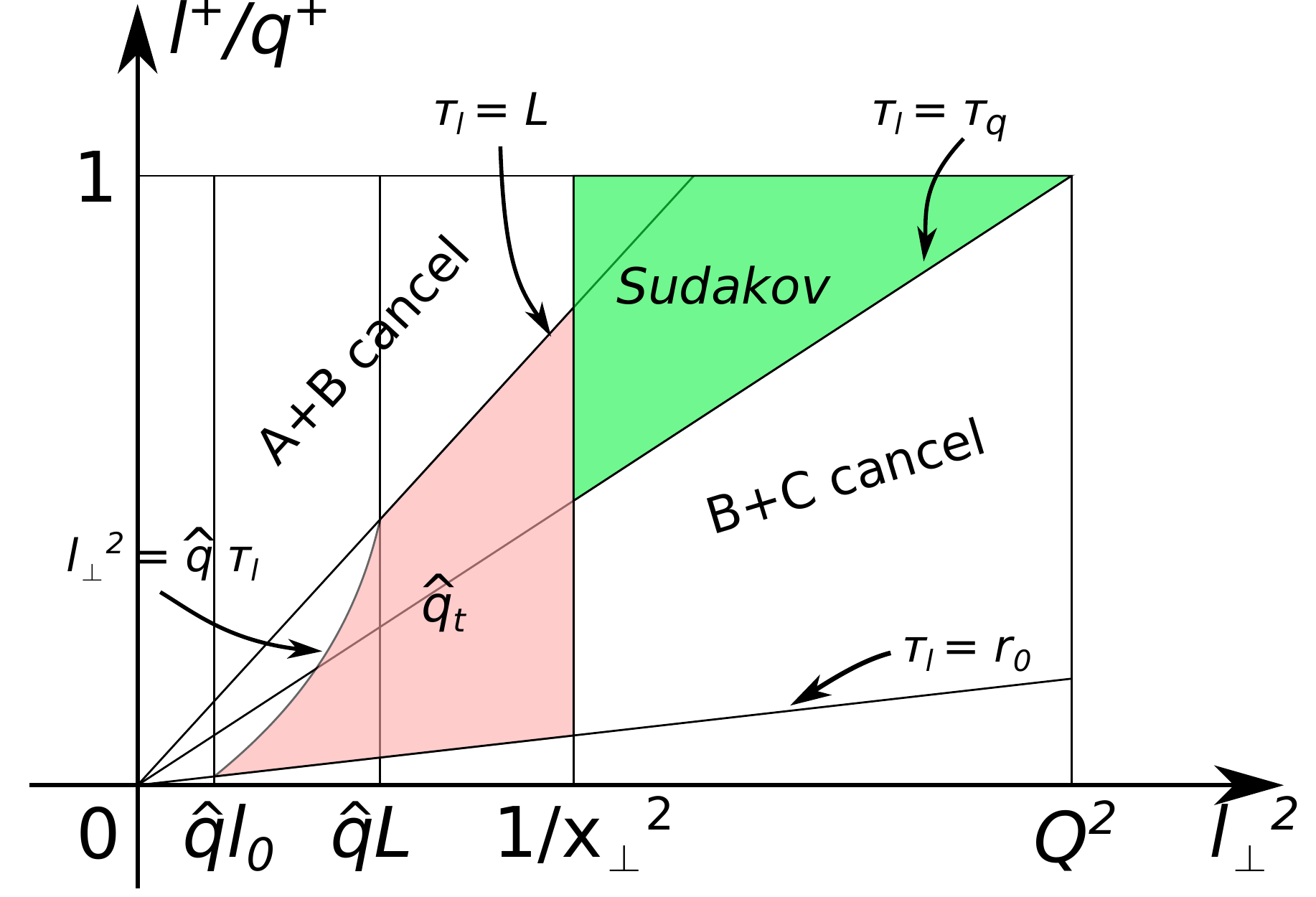}\hspace{1cm}\includegraphics[width=8cm]{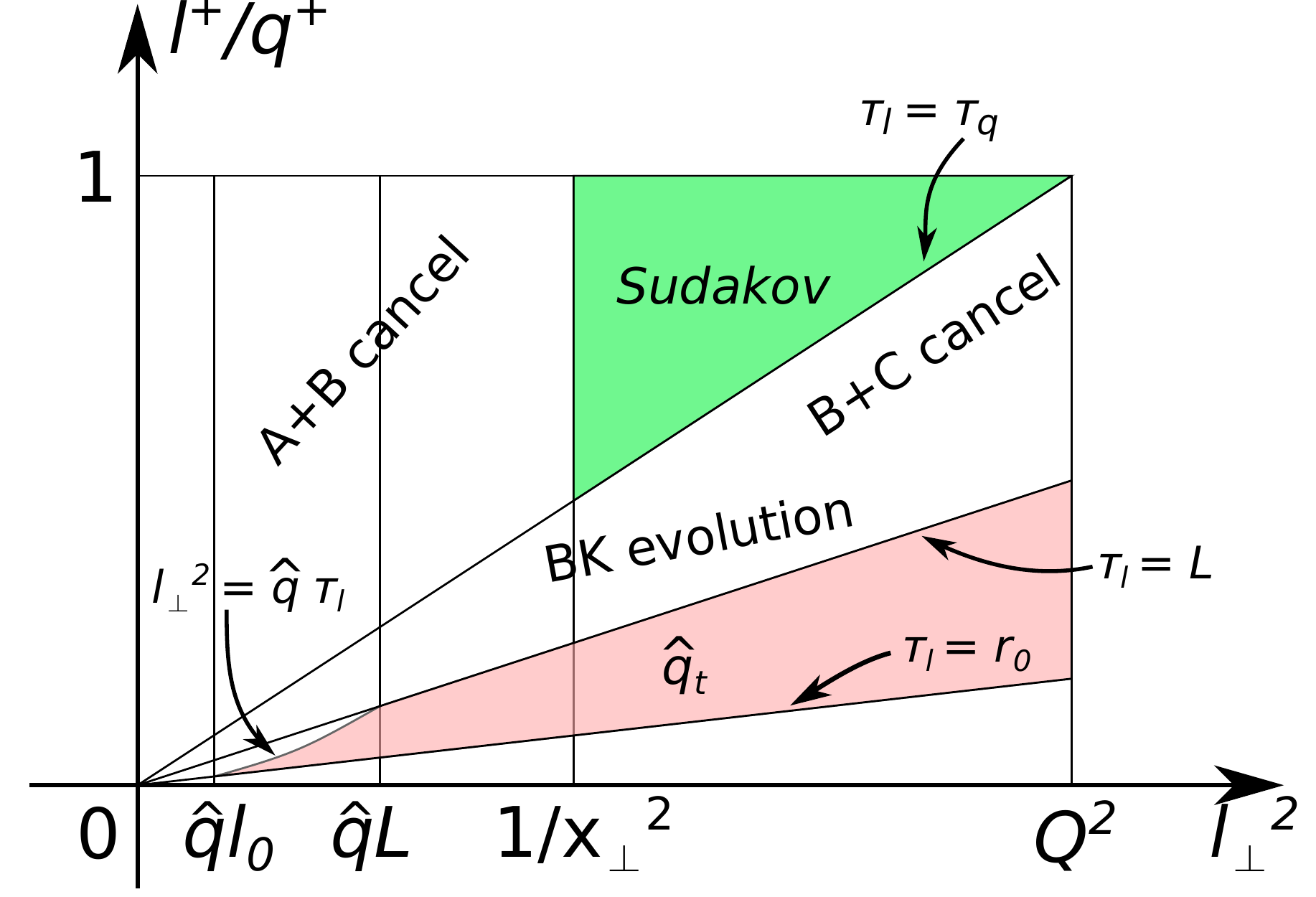}
\end{center}
\caption[*]{Left figure corresponds to large medium case with $\tau_q\equiv 2q_+/Q^2 \ll L$: The red shaded region shows when in-medium radiation contributes to transverse momentum broadening. The cancellation of the double logarithmic contributions from Fig.~\ref{fig:nlo} is indicated. The Sudakov contribution, coming from graph B in Fig.~\ref{fig:nlo}, is shown as the green shaded region in the upper right hand corner. Here the formation time $\tau_l\equiv \frac{2l^+}{l_\perp^2}$. Right figure indicates the shock wave limit with $\tau_q \gg L$: In-medium radiation and multiple scattering, shown in the shaded region, now gives the initial condition, at $\tau_l=L$ or $Y_0=\ln LM$ with $M$ the nucleon mass, for BK evolution. The region where Sudakov suppression effects come from is indicated in the upper right hand part of the figure. Regions where $A+B$ of Fig.~\ref{fig:nlo} and $B+C$ of Fig.~\ref{fig:nlo} cancel are shown. Above the line $\tau_l=L$ and below the line $\tau_l=\tau_q$ is the region of BK evolution. }
\label{qm2}
\end{figure}

Through the one-loop calculation of quark jet production in DIS on a large nucleus by allowing one extra gluon radiation, we integrate over the full phase space of the radiated gluon, and show that the medium induced broadening effects can be separated from the conventional Sudakov effects coming from parton showers in the vacuum in both large medium and shock wave cases, which are summarized in the left and right phase space plots in Fig.~\ref{qm2}, respectively. As shown in Fig.~\ref{qm2}, the transverse momentum broadening due to Sudakov effects and medium induced radiation as well as the small-$x$ evolution (the small-$x$ evolution is absent in the former case since $x_{\textrm{Bj}}$ is taken to be large) come from different regions of the phase space of the radiated gluon. 

A similar calculation for the process of producing a gluon jet in DIS with a gluonic current\cite{Kovchegov:1998bi, Mueller:1999wm} can also be performed, which leads to a similar result as the quark jet production considered in this paper. In this case, the color factor of the Sudakov double logarithm is $C_A$ instead of $C_F$, and the medium effects is taken care of by the WW gluon distribution with the possible corresponding small-$x$ evolution\cite{Dominguez:2011gc} and the gluonic $\hat q$ in the initial condition. Based on these examples, we argue that there should be a factorization between the medium $P_T$ broadening effects and the Sudakov effects in general for hard processes, since the Sudakov effects normally arise from vacuum soft-collinear gluon radiation and they only depend on the virtuality of the considered hard processes and the measured transverse momentum. 

The separation of Sudakov effects and medium broadening effects allows us to have a more sophisticated framework to compute the medium $P_T$ broadening effects in hard processes especially the dijet or dihadron productions in heavy ion collisions where Sudakov effects are not negligible. We can further use these processes as probes to quantitatively extract the values of $\hat q$ at RHIC and the LHC energies (see e.g., Refs.~\cite{ Mueller:2016gko, Chen:2016vem}).

\begin{acknowledgments}
This work was supported in part by the U.S. Department of Energy under the contracts DE-AC02-05CH11231 and DE-FG02-92ER40699, and by the NSFC under Grant No.~11575070 and No.~11521064. This material is also based upon work supported by the U.S. Department of Energy, Office of Science, Office of Nuclear Physics under Award Number DE-SC0004286 (BW). B. W. would like to thank Yuri Kovchegov for useful and informative discussions. Three of the authors, A. H. M, B. W. and B. X., would like to thank Dr. Jian-Wei Qiu and the nuclear theory group at BNL for the hospitality and support during their visit when this work was finalized.
\end{acknowledgments}

\appendix

\section{The Sudakov double log in DIS}

\begin{figure}[tbp]
\begin{center}
\includegraphics[width=18cm]{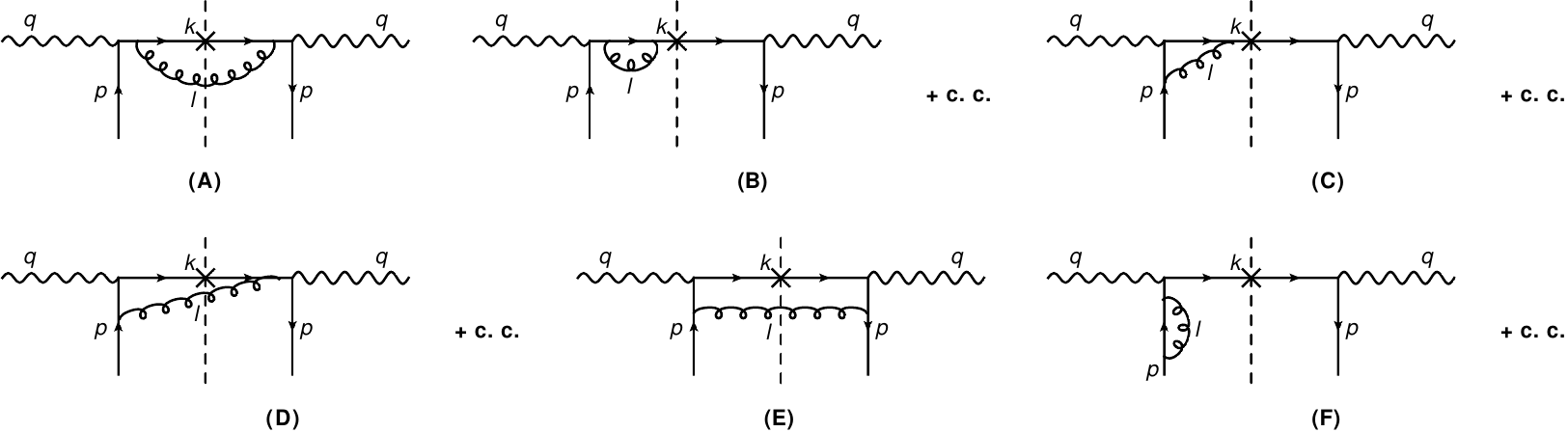}
\end{center}
\caption[*]{Diagrams for DIS at the next-to-leading order (NLO).}
\label{fig:nlo}
\end{figure}

We shall carry out the calculation by choosing a frame in which the nucleus is moving along the negative $z$-direction with momentum $P^\mu$. At leading-order, the virtual photon knocks out a quark carrying an energy fraction $z=\frac{p^-}{P^-}$. The overall normalization is chosen to give
\begin{eqnarray}
\begin{array}{l}
\includegraphics[width=4cm]{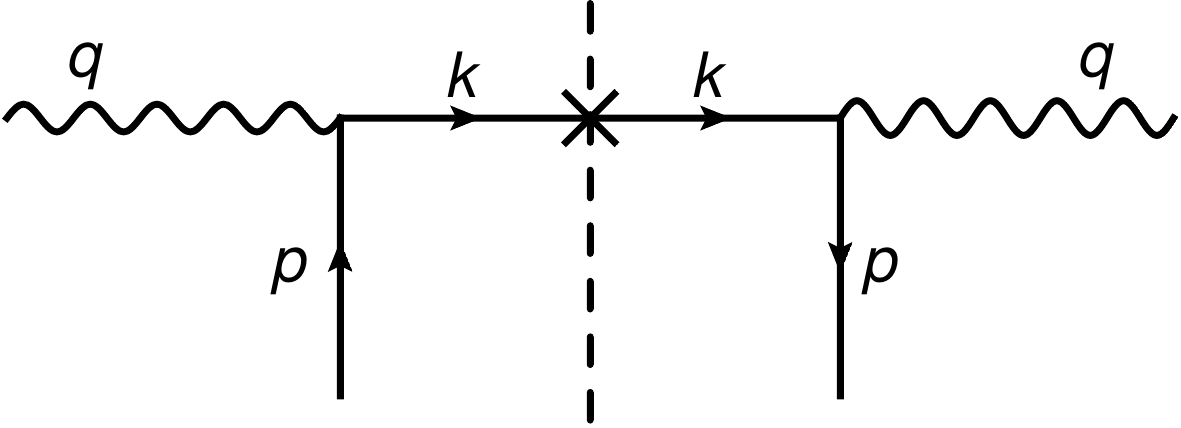}
\end{array}
&&=\int\frac{dk^+ dk^-}{(2\pi)^2}\int\frac{d^2p_\perp}{2(2\pi)^3}\int dz \bar u_p\slashed{\epsilon}^*_{T\lambda} \slashed{k} \slashed{\epsilon}_{T\lambda} u_p 2\pi\delta(k^2) (2\pi)^4 \delta(q+p-k) q_T(z,p_\perp)\nonumber\\
&&=\frac{1}{2}\int dz \bar u_p\slashed{\epsilon}^*_{T\lambda} (\slashed{q}+\slashed{p}) \slashed{\epsilon}_{T\lambda} u_p\delta((q+p)^2) q_T(z,k_\perp)\nonumber\\
&&\approx\frac{1}{2}\int dz \bar u_p\slashed{\epsilon}^*_{T\lambda} q^+ \gamma^- \slashed{\epsilon}_{T\lambda} u_p\delta((q+p)^2) q_T(z,k_\perp)=x_{Bj} q_T(x_{Bj},k_\perp),
\end{eqnarray}
where $\epsilon_{T\lambda}^\mu$ is the transverse polarization vector, $x_{Bj}\equiv Q^2/(2q\cdot p)$ and $q_T(z,k_\perp)$ is the unintegrated (TMD) quark distribution. As illustrated in Fig. \ref{fig:nlo}, there are 6 diagrams at NLO. We shall choose $A^+=0$ lightcone gauge. For the double log result, the gluon in these diagrams can be taken as soft, that is, $l^+\ll q^+\approx k^+$ and $l^-\ll p^-$. In this gauge (E) and (F) do not contribute. In terms of
\begin{eqnarray}
P_{\alpha\beta}(k)=-g_{\alpha\beta}+\frac{k_\alpha\eta_\beta+k_\beta\eta_\alpha}{k^+}\qquad\text{with $\eta_\alpha=g_\alpha^+$},
\end{eqnarray}
one has
\begin{eqnarray}
A=\frac{g^2 C_F}{2}\int_0^1 dz\int \frac{d^2l_\perp dl^+}{(2\pi)^3 2 l^+} \frac{\bar u_p\slashed{\epsilon}^*_{T\lambda} (\slashed{k}+\slashed{l}) \gamma^\alpha\slashed{k}\gamma^\beta(\slashed{k}+\slashed{l}) \slashed{\epsilon}_{T\lambda} u_p}{[(k+l)^2+i\epsilon]^2} P_{\alpha\beta}(l)\delta((q+p-l)^2) q_T(z,k_\perp + l_\perp).
\end{eqnarray}
By keeping the leading order in $q^+$, one has
\begin{eqnarray}
A&&\approx\frac{g^2 C_F}{4P^-}\int dz\int \frac{d^2l_\perp dl^+}{(2\pi)^3 2 l^+}\frac{1}{4(l^-)^2} [\bar u_p\slashed{\epsilon}^*_{T\lambda} \gamma^- \gamma^+\gamma^-\gamma^+\gamma^- \slashed{\epsilon}_{T\lambda} u_p] P_{++}(l)\delta\left(z-\frac{l^-}{P^-}-x_{Bj}\right) q_T(z,k_\perp + l_\perp)\nonumber\\
&&=\left.\frac{\alpha_s C_F}{\pi}\int \frac{dl_\perp^2}{l_\perp^2} \frac{dl^+}{l^+} z q_T(z,k_\perp+l_\perp)\right|_{z=x_{Bj}+\frac{l^-}{P^-}}=\left.\frac{\alpha_s C_F}{\pi}\int_0^{Q^2} \frac{dl_\perp^2}{l_\perp^2}\int_0^1\frac{d(1-\xi)}{1-\xi} z q_T(z,k_\perp+l_\perp)\right|_{z=x_{Bj}+\frac{l^-}{P^-}}.
\end{eqnarray}
Here, the double log region lies only in the range $l_\perp^2 \lesssim k_\perp^2$. Diagram (B) can be easily obtained from the conservation of probability, that is,
\begin{eqnarray}
B=-\frac{\alpha_s C_F}{\pi}\int_0^{Q^2} \frac{dl_\perp^2}{l_\perp^2}\int_0^1\frac{d(1-\xi)}{1-\xi} x_{Bj} q_T(x_{Bj},k_\perp).
\end{eqnarray}
Since $\frac{l^-}{P^-}\ll 1$, one can neglect it compared to $x_{Bj}$ and one has
\begin{eqnarray}
A+B\approx-\frac{\alpha_s C_F}{\pi}\int_{k_\perp^2}^{Q^2} \frac{dl_\perp^2}{l_\perp^2}\int_0^1\frac{d(1-\xi)}{1-\xi} x_{Bj} q_T(x_{Bj},k_\perp).
\end{eqnarray}

Diagram (C) is given by
\begin{eqnarray}
C=i g^2 C_F\int dz\int \frac{d^4l}{(2\pi)^4} \frac{\bar u_p\slashed{\epsilon}^*_{T\lambda} \slashed{k} \gamma^\alpha (\slashed{k}-\slashed{l})\slashed{\epsilon}_{T\lambda}(\slashed{p}-\slashed{l}) \gamma^\beta u_p}{(l^2+i\epsilon)[(k-l)^2+i\epsilon] [(p-l)^2+i\epsilon]} P_{\alpha\beta}(l)\delta((q+p)^2) q_T(z,k_\perp).
\end{eqnarray}
As before, $q^+\approx k^+$ is taken to be large and, as a result, one has
\begin{eqnarray}
C\approx \frac{i g^2 C_F}{P^-} q_T(x_{Bj}, k_\perp)\int \frac{d^4l}{(2\pi)^4} \frac{\bar u_p\gamma^- (\slashed{p}-\slashed{l}) \gamma^\beta u_p}{(l^2+i\epsilon)[-l^-+i\epsilon] [(p-l)^2+i\epsilon]} P_{+\beta}(l).
\end{eqnarray}
The integrand of the above integral has a pole given by
\begin{eqnarray}
0=(p-l)^2+i\epsilon,
\end{eqnarray}
that is,
\begin{eqnarray}
l^-\approx\frac{l_\perp^2-i\epsilon}{2l^+}+p^-\qquad\text{with$\qquad l_\perp>p_\perp$}.
\end{eqnarray}
By using this fact, we obtain
\begin{eqnarray}
C&&\approx\frac{i g^2 C_F}{P^-} q_T(x_{Bj}, k_\perp)\int \frac{d^2l_\perp dl^+ dl^-}{(2\pi)^4} \frac{\bar u_p\gamma^- (l_\perp\cdot \gamma_\perp) \gamma_{\perp}^i u_p}{(l^2+i\epsilon)[-l^-+i\epsilon] [(p-l)^2+i\epsilon]} P_{+i}(l)\nonumber\\
&&=\frac{g^2 C_F}{P^-} q_T(x_{Bj}, k_\perp)\int\limits_{l^+>0} \frac{d^2l_\perp dl^+}{(2\pi)^3} \frac{\bar u_p\gamma^- l_\perp\cdot \gamma_\perp l_\perp\cdot \gamma_\perp u_p}{l_\perp^2 (- 2 l^+ p^- - l_\perp^2) l^+}.
\end{eqnarray}
Since $p^-\approx -q^-$, one has
\begin{eqnarray}
C\approx \frac{g^2 C_F}{P^-} q_T(x_{Bj}, k_\perp)\int\limits_{l^->q^-}\frac{d^2l_\perp dl^+}{(2\pi)^3} \frac{\bar u_p\gamma^- u_p}{l_\perp^2 l^+}=\frac{\alpha_s C_F}{\pi}x_{Bj} q_T(x_{Bj},k_\perp)\int_{0}^{Q^2} \frac{dl_\perp^2}{l_\perp^2}\int_0^{\frac{l_\perp^2}{Q^2}}\frac{d(1-\xi)}{1-\xi}.
\end{eqnarray}
Similarly, including the contribution from (D) gives
\begin{eqnarray}
C+D=\frac{\alpha_s C_F}{\pi}x_{Bj} q_T(x_{Bj},k_\perp)\int_{k_\perp^2}^{Q^2} \frac{dl_\perp^2}{l_\perp^2}\int_{0}^{\frac{l_\perp^2}{Q^2}}\frac{d(1-\xi)}{1-\xi} .
\end{eqnarray}
At the end, the Sudakov double log at NLO is given by
\begin{eqnarray}
A+B+C+D=-\frac{\alpha_s C_F}{2\pi} x_{Bj} q_T(x_{Bj},k_\perp) \ln^2\frac{Q^2}{k_\perp^2} .
\end{eqnarray}
In the above calculation, we have neglected transverse momentum conservation when gluons are emitted. It is straight forward to repeat the above calculation in transverse coordinate space in order to restore transverse momentum conservation for arbitrary number of gluon emission. In this case, it is convenient to combine the amplitudes and conjugate amplitudes of the diagrams in Fig. \ref{fig:nlo} into a dipole-like picture\cite{Zakharov:1996fv,Liou:2013qya}. Then, the inverse of the dipole size $x_\perp\sim \frac{1}{k_\perp}$ plays the same role as $k_\perp$ in the above discussion. Therefore, one arrives at
\begin{eqnarray}
A+B+C+D=-\frac{\alpha_s C_F}{2\pi} x_{Bj} q_T(x_{Bj},1/x_\perp) \ln^2 Q^2 x_\perp^2
\end{eqnarray}
with $x_{Bj} q_T(x_{Bj},1/x_\perp)$ the quark distribution in the coordinate space.


\begin{thebibliography}{9}

%
\bibitem{Gyulassy:2004zy} 
  M.~Gyulassy and L.~McLerran,
  Nucl.\ Phys.\ A {\bf 750}, 30 (2005)
  [nucl-th/0405013].
  
  
    %
\bibitem{Gyulassy:1993hr} 
  M.~Gyulassy and X.~n.~Wang,
  Nucl.\ Phys.\ B {\bf 420}, 583 (1994)
  [nucl-th/9306003].
  
%
\bibitem{Baier:1996kr} 
  R.~Baier, Y.~L.~Dokshitzer, A.~H.~Mueller, S.~Peigne and D.~Schiff,
  Nucl.\ Phys.\ B {\bf 483}, 291 (1997)
  [hep-ph/9607355].
  

\bibitem{Baier:1996sk} 
  R.~Baier, Y.~L.~Dokshitzer, A.~H.~Mueller, S.~Peigne and D.~Schiff,
  Nucl.\ Phys.\ B {\bf 484}, 265 (1997)
  [hep-ph/9608322].
  
\bibitem{Baier:1998kq} 
  R.~Baier, Y.~L.~Dokshitzer, A.~H.~Mueller and D.~Schiff,
  Nucl.\ Phys.\ B {\bf 531}, 403 (1998)
  [hep-ph/9804212].
  
  
\bibitem{Zakharov:1996fv} 
  B.~G.~Zakharov,
  JETP Lett.\  {\bf 63}, 952 (1996)
  [hep-ph/9607440]; 
  JETP Lett.\  {\bf 65}, 615 (1997)
  [hep-ph/9704255].



  
  
%
\bibitem{Adams:2003kv} 
  J.~Adams {\it et al.} [STAR Collaboration],
  Phys.\ Rev.\ Lett.\  {\bf 91}, 172302 (2003)
  [nucl-ex/0305015].
  
\bibitem{Adams:2003im} 
  J.~Adams {\it et al.} [STAR Collaboration],
  Phys.\ Rev.\ Lett.\  {\bf 91}, 072304 (2003)
  [nucl-ex/0306024].
  
\bibitem{Adler:2003qi} 
  S.~S.~Adler {\it et al.} [PHENIX Collaboration],
  Phys.\ Rev.\ Lett.\  {\bf 91}, 072301 (2003)
  [nucl-ex/0304022].

%
\bibitem{Adler:2002tq} 
  C.~Adler {\it et al.} [STAR Collaboration],
  Phys.\ Rev.\ Lett.\  {\bf 90}, 082302 (2003)
  [nucl-ex/0210033].

  

  
  %
\bibitem{CMS:2012aa} 
  S.~Chatrchyan {\it et al.} [CMS Collaboration],
  Eur.\ Phys.\ J.\ C {\bf 72}, 1945 (2012)
  [arXiv:1202.2554 [nucl-ex]].
  
\bibitem{Abelev:2012hxa} 
  B.~Abelev {\it et al.} [ALICE Collaboration],
  Phys.\ Lett.\ B {\bf 720}, 52 (2013)
  [arXiv:1208.2711 [hep-ex]].


%
\bibitem{Aad:2012vca} 
  G.~Aad {\it et al.} [ATLAS Collaboration],
  Phys.\ Lett.\ B {\bf 719}, 220 (2013)
  [arXiv:1208.1967 [hep-ex]].
  
  %
\bibitem{Aamodt:2011vg} 
  K.~Aamodt {\it et al.} [ALICE Collaboration],
  Phys.\ Rev.\ Lett.\  {\bf 108}, 092301 (2012)
  [arXiv:1110.0121 [nucl-ex]].
  
\bibitem{Chatrchyan:2011sx} 
  S.~Chatrchyan {\it et al.} [CMS Collaboration],
  Phys.\ Rev.\ C {\bf 84}, 024906 (2011)
  [arXiv:1102.1957 [nucl-ex]].
  
\bibitem{Aad:2010bu} 
  G.~Aad {\it et al.} [ATLAS Collaboration],
  Phys.\ Rev.\ Lett.\  {\bf 105}, 252303 (2010)
  [arXiv:1011.6182 [hep-ex]].
  
  

    
   %
\bibitem{Qin:2010mn} 
  G.~Y.~Qin and B.~Muller,
  Phys.\ Rev.\ Lett.\  {\bf 106}, 162302 (2011)
  [Phys.\ Rev.\ Lett.\  {\bf 108}, 189904 (2012)]
  [arXiv:1012.5280 [hep-ph]].
  
  
 \bibitem{Collins:1984kg}
  J.~C.~Collins, D.~E.~Soper and G.~F.~Sterman,
  Nucl.\ Phys.\ B {\bf 250}, 199 (1985).
  
  %
\bibitem{Banfi:2008qs} 
  A.~Banfi, M.~Dasgupta and Y.~Delenda,
  Phys.\ Lett.\ B {\bf 665}, 86 (2008)
  [arXiv:0804.3786 [hep-ph]].
  
  
  \bibitem{Mueller:2012uf} 
  A.~H.~Mueller, B.~-W.~Xiao and F.~Yuan,
  Phys.\ Rev.\ Lett.\  {\bf 110}, 082301 (2013)
  [arXiv:1210.5792 [hep-ph]].  

    %
\bibitem{Mueller:2013wwa}
  A.~H.~Mueller, B.~-W.~Xiao and F.~Yuan,
  Phys.\ Rev.\ D {\bf 88}, 114010 (2013).


\bibitem{Sun:2014gfa} 
  P.~Sun, C.-P.~Yuan and F.~Yuan,
  Phys.\ Rev.\ Lett.\  {\bf 113}, no. 23, 232001 (2014);
  P.~Sun, C.-P.~Yuan and F.~Yuan,
  arXiv:1506.06170 [hep-ph].

  %
\bibitem{Abazov:2004hm}
  V.~M.~Abazov {\it et al.}  [D0 Collaboration],
  Phys.\ Rev.\ Lett.\  {\bf 94}, 221801 (2005).
  
  

%
\bibitem{Khachatryan:2011zj}
  V.~Khachatryan {\it et al.}  [CMS Collaboration],
  Phys.\ Rev.\ Lett.\  {\bf 106}, 122003 (2011).

%
\bibitem{Chatrchyan:2014hqa} 
  S.~Chatrchyan {\it et al.} [CMS Collaboration],
  Eur.\ Phys.\ J.\ C {\bf 74}, no. 7, 2951 (2014)
  [arXiv:1401.4433 [nucl-ex]].
  
  
\bibitem{Mueller:2016gko} 
  A.~H.~Mueller, B.~Wu, B.~W.~Xiao and F.~Yuan,
  arXiv:1604.04250 [hep-ph].
  
 
\bibitem{Chen:2016vem} 
  L.~Chen, G.~Y.~Qin, S.~Y.~Wei, B.~W.~Xiao and H.~Z.~Zhang,
  arXiv:1607.01932 [hep-ph].
  
  
\bibitem{Luo:1993ui} 
  M.~Luo, J.~w.~Qiu and G.~F.~Sterman,
  Phys.\ Rev.\ D {\bf 49}, 4493 (1994).
  
  
\bibitem{Zhang:2014dya} 
  J.~J.~Zhang, J.~H.~Gao and X.~N.~Wang,
  Phys.\ Rev.\ D {\bf 91}, no. 1, 014026 (2015)
  [arXiv:1411.5435 [hep-ph]].
  
\bibitem{Kang:2016ron} 
  Z.~B.~Kang, J.~W.~Qiu, X.~N.~Wang and H.~Xing,
  arXiv:1605.07175 [hep-ph].
  
\bibitem{Kovchegov:1996ty} 
  Y.~V.~Kovchegov,
  Phys.\ Rev.\ D {\bf 54}, 5463 (1996)
  [hep-ph/9605446].
  
  \bibitem{Kovchegov:1998bi} 
  Y.~V.~Kovchegov and A.~H.~Mueller,
  Nucl.\ Phys.\ B {\bf 529}, 451 (1998)
  [hep-ph/9802440].

\bibitem{Kharzeev:2003wz} 
  D.~Kharzeev, Y.~V.~Kovchegov and K.~Tuchin,
  Phys.\ Rev.\ D {\bf 68}, 094013 (2003)
  [hep-ph/0307037].

%
\bibitem{Kovchegov:2001sc} 
  Y.~V.~Kovchegov and K.~Tuchin,
  Phys.\ Rev.\ D {\bf 65}, 074026 (2002)
  [hep-ph/0111362].
  
  
\bibitem{Mueller:2012bn} 
  A.~H.~Mueller and S.~Munier,
  Nucl.\ Phys.\ A {\bf 893}, 43 (2012)
  [arXiv:1206.1333 [hep-ph]].

 \bibitem{Liou:2013qya} 
  T.~Liou, A.~H.~Mueller and B.~Wu,
  Nucl.\ Phys.\ A {\bf 916}, 102 (2013)
  [arXiv:1304.7677 [hep-ph]].
  
   \bibitem{Wu:2011kc} 
  B.~Wu,
  JHEP {\bf 1110}, 029 (2011)
  [arXiv:1102.0388 [hep-ph]].
  
  
  
  


  
\bibitem{Blaizot:2014bha} 
  J.~P.~Blaizot and Y.~Mehtar-Tani,
  Nucl.\ Phys.\ A {\bf 929}, 202 (2014)
  [arXiv:1403.2323 [hep-ph]].
  
  %
\bibitem{Iancu:2014kga} 
  E.~Iancu,
  JHEP {\bf 1410}, 95 (2014)
  [arXiv:1403.1996 [hep-ph]].
  



  %
\bibitem{Iancu:2014sha} 
  E.~Iancu and D.~N.~Triantafyllopoulos,
  Phys.\ Rev.\ D {\bf 90}, no. 7, 074002 (2014)
  [arXiv:1405.3525 [hep-ph]].
  
  %
\bibitem{Gorshkov:1966ht} 
  V.~G.~Gorshkov, V.~N.~Gribov, L.~N.~Lipatov and G.~V.~Frolov,
  Sov.\ J.\ Nucl.\ Phys.\  {\bf 6}, 95 (1968)
  [Yad.\ Fiz.\  {\bf 6}, 129 (1967)].
  
  \bibitem{Kovchegov:2012mbw} 
  Y.~V.~Kovchegov and E.~Levin,
  ``Quantum chromodynamics at high energy,'' (2012), Cambridge University Press. 

   
   \bibitem{Mueller:1999wm}  A.~H.~Mueller,  
Nucl.\ Phys.\ B \textbf{558}, 285 (1999)  [hep-ph/9904404].  

   

\bibitem{Kwiecinski:2002ep} 
  J.~Kwiecinski and A.~M.~Stasto,
  Phys.\ Rev.\ D {\bf 66}, 014013 (2002)
  [hep-ph/0203030].
  
  %
\bibitem{Iancu:2002tr} 
  E.~Iancu, K.~Itakura and L.~McLerran,
  Nucl.\ Phys.\ A {\bf 708}, 327 (2002)
  [hep-ph/0203137].
  
  %
\bibitem{Balitsky:1995ub} 
  I.~Balitsky,
  Nucl.\ Phys.\ B {\bf 463}, 99 (1996)
  [hep-ph/9509348].
  
\bibitem{Kovchegov:1999yj} 
  Y.~V.~Kovchegov,
  Phys.\ Rev.\ D {\bf 60}, 034008 (1999)
  [hep-ph/9901281].
  
\bibitem{JalilianMarian:1997gr} 
  J.~Jalilian-Marian, A.~Kovner, A.~Leonidov and H.~Weigert,
  Phys.\ Rev.\ D {\bf 59}, 014014 (1998)
  [hep-ph/9706377].
  
\bibitem{Iancu:2001ad} 
  E.~Iancu, A.~Leonidov and L.~D.~McLerran,
  Phys.\ Lett.\ B {\bf 510}, 133 (2001)
  [hep-ph/0102009].

  %
\bibitem{Dominguez:2011gc} 
  F.~Dominguez, A.~H.~Mueller, S.~Munier and B.~W.~Xiao,
  Phys.\ Lett.\ B {\bf 705}, 106 (2011)
  [arXiv:1108.1752 [hep-ph]].

 

\end{thebibliography}
\end{document}